\documentclass[twocolumn]{aastex63}
\usepackage{scrextend}
\deffootnote[0.45em]{0.2em}{0.2em}{\textsuperscript{\thefootnotemark}\,}
\usepackage{courier}
\usepackage[T1]{fontenc}
\usepackage{mathtools}
\usepackage{changepage} 
\usepackage{url,multirow}
\usepackage{appendix}
\usepackage{microtype}
\DeclareUnicodeCharacter{2212}{-}
\usepackage{lineno}
\usepackage{textcomp}
\usepackage{gensymb}
\usepackage{todonotes}
\usepackage{lipsum}

\title{2FHL~J1745.1--3035: A Newly Discovered, Powerful Pulsar Wind Nebula Candidate}

\def\xmm{{XMM-{\it Newton\/}}}

\def\cha{{\it Chandra}}

\def\nus{{\it NuSTAR}}
\def\nhlos{N$_{\rm H,los}$}

\def\hess{{H.E.S.S.}}

\def\g-rays{{$\gamma$ rays}}
\def\lat{{\it Fermi}--LAT}
\def\nhlos{N$_{\rm H,los}$}

\def\aap{A\&A}

\def\apjl{ApJL}
\def\apjs{ApJS}

\shorttitle{2FHL~J1745.1--3035: A Newly Discovered, Powerful Pulsar Wind Nebula}
\shortauthors{Marchesi et al.}

\begin{document}

\title{2FHL~J1745.1--3035: A Newly Discovered, Powerful Pulsar Wind Nebula Candidate}

\author[0000-0001-5544-0749]{S. Marchesi}
\affiliation{Dipartimento di Fisica e Astronomia (DIFA), Università di Bologna, via Gobetti 93/2, I-40129 Bologna, Italy}
\affiliation{Department of Physics and Astronomy, Clemson University, Kinard Lab of Physics, Clemson, SC 29634-0978, USA}
\affiliation{INAF - Osservatorio di Astrofisica e Scienza dello Spazio di Bologna, Via Piero Gobetti, 93/3, 40129, Bologna, Italy}

\author[0000-0001-9633-3165]{J. Eagle}
\affiliation{NASA Goddard Space Flight Center, Greenbelt, MD 20771, USA}

\author[0000-0002-6584-1703]{M.~Ajello}
\affiliation{Department of Physics and Astronomy, Clemson University, Kinard Lab of Physics, Clemson, SC 29634-0978, USA}

\author[0000-0002-0394-3173]{D. Castro}
\affiliation{Harvard-Smithsonian Center for Astrophysics, Cambridge, MA 02138, USA}

\author[0000-0002-3433-4610]{A.~Dom\'inguez}
\affiliation{IPARCOS and Department of EMFTEL, Universidad Complutense de Madrid, E-28040 Madrid, Spain}

\author[0000-0002-9709-5389]{K.~Mori}
\affiliation{Columbia Astrophysics Laboratory, Columbia University, New York, NY 10027, USA}

\author[0000-0001-7523-570X]{L.~Tibaldo}
\affiliation{IRAP, Universit\'{e} de Toulouse, CNRS, CNES, UPS, 9 Avenue Colonel Roche, 31028 Toulouse, Cedex 4, France}

\author[0000-0001-5506-9855]{J.~Tomsick}
\affiliation{Space Sciences Laboratory, University of California, Berkeley, 7 Gauss Way, Berkeley, CA 94720-7450, USA}

\author[0000-0003-1006-924X]{A.~Traina}
\affiliation{INAF - Osservatorio di Astrofisica e Scienza dello Spazio di Bologna, Via Piero Gobetti, 93/3, 40129, Bologna, Italy}
\affiliation{Dipartimento di Fisica e Astronomia (DIFA), Università di Bologna, via Gobetti 93/2, I-40129 Bologna, Italy}

\author[0000-0002-8853-9611]{C.~Vignali}
\affiliation{Dipartimento di Fisica e Astronomia (DIFA), Università di Bologna, via Gobetti 93/2, I-40129 Bologna, Italy}
\affiliation{INAF - Osservatorio di Astrofisica e Scienza dello Spazio di Bologna, Via Piero Gobetti, 93/3, 40129, Bologna, Italy}

\author[0000-0001-6320-1801]{R.~Zanin}\affiliation{Cherenkov Telescope Array Observatory gGmbH, Via Piero Gobetti, 93/3, 40129, Bologna, Italy}

\begin{abstract}
We present a multi-epoch, multi-observatory X-ray analysis for 2FHL~J1745.1--3035, a newly discovered very high energy Galactic source detected by the \textit{Fermi} Large Area Telescope (LAT) located in close proximity to the Galactic Center (\textit{l}=358.5319$^\circ$; \textit{b}=--0.7760$^\circ$). The source shows a very hard $\gamma$-ray photon index above 50\,GeV, $\Gamma_\gamma$=1.2$\pm$0.4, and is found to be a TeV-emitter by the \lat.
We conduct a joint \xmm, \cha\, and \nus\ observing campaign, combining archival \xmm\ observations, to study the X-ray spectral properties of 2FHL~J1745.1--3035 over a time-span of over 20 years. The joint X-ray spectrum is best-fitted as a broken power law model with break energy E$_{b}\sim$7\,keV: the source is very hard at energies below 10\,keV, with $\Gamma_{\rm 1}\sim$0.6, and significantly softer in the higher energy range measured by \nus\ with $\Gamma_{\rm 2}\sim$1.9.
We also perform a spatially resolved X-ray analysis with \cha, finding evidence for marginal extension (up to an angular size $r\sim$5$^{\prime\prime}$), a result that supports a compact pulsar wind nebula scenario.
Based on the X-ray and $\gamma$-ray properties, 2FHL~J1745.1--3035 is a powerful pulsar wind nebula candidate. Given its nature as an extreme TeV emitter, further supported by the detection of a coincident TeV extended source HESS J1745-303, 2FHL~J1745.1--3035 is an ideal candidate for a follow-up with the upcoming Cherenkov Telescope Array.

\end{abstract}

\section{Introduction} \label{sec:intro}
The Galactic plane is rich with efficient accelerators producing cosmic rays (CRs, both leptons and hadrons),  neutrinos (via hadronic interactions), and  energetic $\gamma$-ray photons.  Very high-energy (VHE, $>$50\,GeV) $\gamma$-rays provide a direct view of some of the most extreme environments in the Galaxy: indeed, $\gamma$-rays represent an excellent probe of non-thermal astrophysical processes, since they are produced by the interaction of relativistic particles. In particular, pulsar wind nebulae (PWNe) are some of the brightest sources in the VHE sky and represent an ideal laboratory for studying phenomena around objects with extreme densities and magnetic field strengths, as well as how relativistic particle winds interact with ambient media. The spinning neutron star, or pulsar, left behind after a supernova (SN) explosion may generate a relativistic wind of electrons and positrons that interact with unheated supernova ejecta, developing a standing shock wave where the electrons and positrons are injected and accelerated into the non-thermal expanding bubble of diffuse plasma, defining the PWN \citep{gaensler_2006,malyshev09}. Since PWNe have well-defined central energy sources (i.e., the pulsar) and many are close enough to be spatially resolved, they allow for studying in great depth both relativistic winds and the shocks that result from these winds colliding with the ambient medium. The radio to X-ray emission of PWNe is caused by synchrotron radiation of relativistic electrons in pulsar winds shocked in the ambient medium \citep{kennel84,kargaltsev08}. Additionally,  the same population of relativistic electrons responsible for the PWN synchrotron emission can scatter off local photon fields, resulting in inverse Compton (IC) emission at $\gamma$-ray energies: notably, PWNe are the dominant class of TeV $\gamma$-ray Galactic sources, as observed by Cherenkov telescopes \citep[for example, 39\,\% of the 31 sources identified by the \hess\ Galactic plane survey are PWNe;][]{hess18}. Therefore, a complete picture of the electron population in PWNe, and a full understanding of the mechanisms underlying VHE  emitters, can be achieved only by constraining simultaneously the synchrotron component, using X-ray observations, and the IC one, with $\gamma$-ray data \citep[e.g.,][]{renaud09,kargaltsev13,eagle22a}. 
 
The {\it Fermi}--LAT 2FHL catalog \citep{ackermann16} contains 12 VHE sources in the Galactic plane ($|b|<10^{\circ}$) having a $\gamma$-ray photon index $\Gamma_\gamma$$<$1.8  and currently lack any association. Such a hard photon index reliably rules out an extragalactic origin for these objects, since at energies $>$50\,GeV the 2FHL blazars generally exhibit a soft spectrum, with an average photon index of $\Gamma_\gamma\sim3.4$. In blazars, this energy range is generally observed as the descending part of the inverse Compton (IC) peak in the spectral energy distribution (SED).  Prior reports have already performed and reported on the analysis for 2 of the 12 unassociated Galactic 2FHL targets. The first study focused on 2FHL~J0826.1--4500, which was found to be a candidate shock-cloud interaction on the western edge of the Vela supernova remnant \citep[SNR;][]{eagle19}. 2FHL~J1703.4--4145 was the second source of the sample analyzed and was similarly found to be the byproduct of a shock-cloud interaction at the edge of a supernova remnant, SNR G344.7--0.1 \citep{eagle20}.

In this work, we present the results of a multi-observatory X-ray monitoring of a third source in the 2FHL VHE subsample, 2FHL~J1745.1--3035. The work is organized as follows: in Section~\ref{sec:2FHLJ1745} we present the \lat\ source properties, and in Section~\ref{sec:analysis} we present the different X-ray datasets available for this target and the data reduction process and analysis performed. In Section~\ref{sec:x-ray_analysis} we present the results of the X-ray spectral fits, both for the single-epoch observations and for the overall, multi-epoch spectrum, and we analyze the \cha\ image of the source, searching for evidence of extended emission. Finally, in Section~\ref{sec:SED_fit} we discuss the properties of 2FHL~J1745.1--3035 by combining the information derived from the X- and $\gamma$-ray data. We summarize the results of the paper in Section~\ref{sec:conclusions}. Through the rest of the work, errors are quoted at the 90\,\% confidence level, unless otherwise stated.

\section{2FHL J1745.1--3035}\label{sec:2FHLJ1745}
\begin{figure} 
 \centering 
\includegraphics[width=0.47\textwidth]{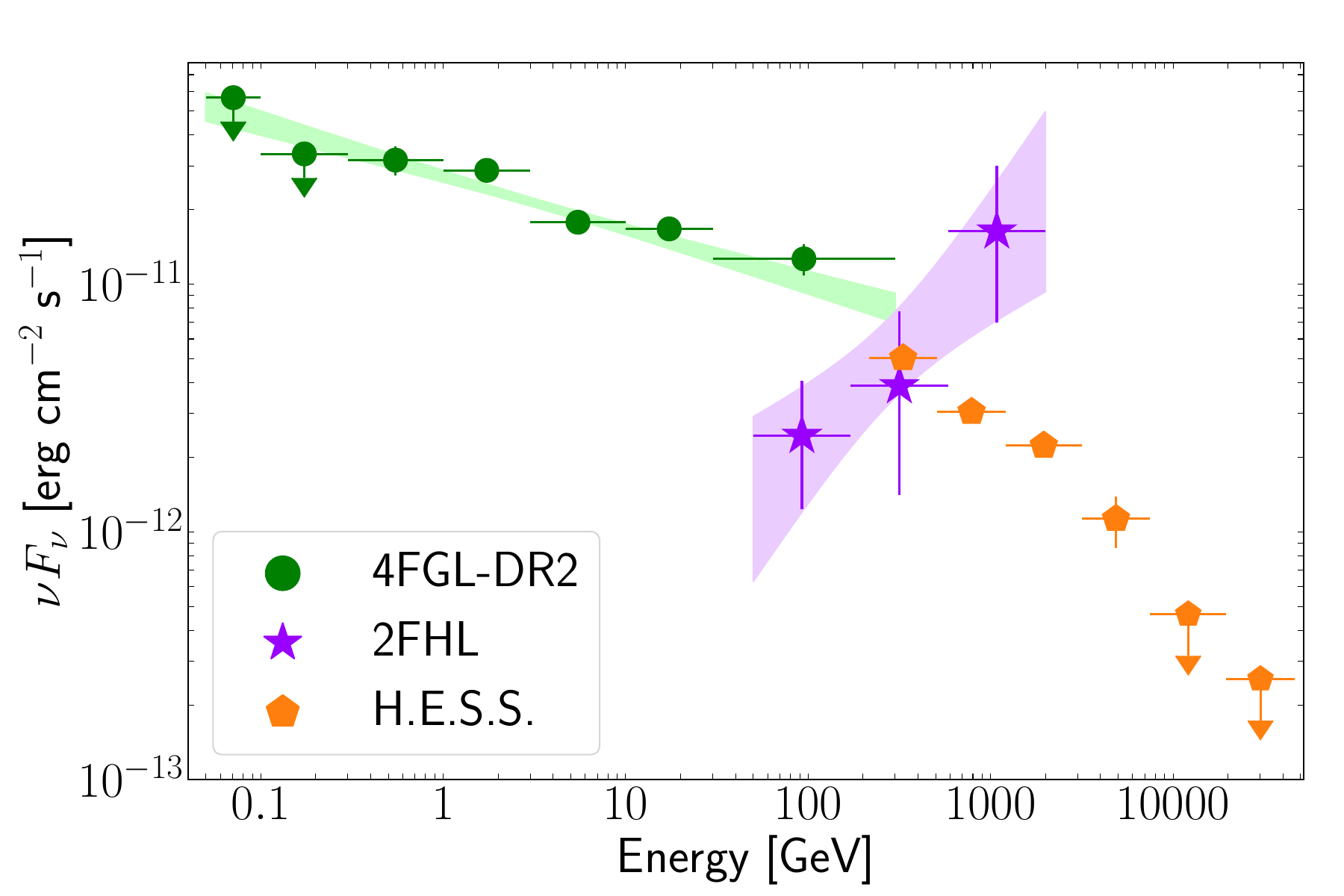} 
\caption{\normalsize 
 \lat\ spectral energy distribution of 2FHL~J1745.1--3035 (violet stars). In the \lat\ data above 50\,GeV, the source has an impressively hard $\gamma$-ray photon index, $\Gamma_\gamma$=1.2$\pm$0.4, and is detected up to $\sim$1\,TeV, making this object a candidate for the most powerful Galactic accelerator. We also plot, for reference, the SEDs for 4FGL J1745.8--3028e (green circles) and HESS~J1745--303 (orange pentagons).
 We discuss in the text the discrepancy between the 2FHL and \hess\ data.
}\label{fig:fermi_SED}
\end{figure}

While inspecting the properties of the new VHE sources reported in the 2FHL catalog, 2FHL~J1745.1--3035 stands out as one of the most intriguing. First of all, the source is located in close proximity to the Galactic Center (\textit{l}=358.5319$^\circ$; \textit{b}=--0.7760$^\circ$). From a $\gamma$-ray perspective, this source is the second brightest\footnote{The brightest being 2FHL~J1703.4--4145, which was already analyzed in \citet{eagle20}.} of the 12 new unassociated VHE sources in the 2FHL sample ($F_{>50~{\rm GeV}}$= (2.69$\pm0.63)$$\times$10$^{-11}$\,erg\,s$^{-1}$\,cm$^{-2}$), and presents a very hard spectrum in the $\gamma$-rays above 50\,GeV (as shown in Figure~\ref{fig:fermi_SED}), having photon index $\Gamma_\gamma$=1.2$\pm0.4$. In fact, \lat\ detected a 2FHL~J1745.1--3035 photon at 940\,GeV, which implies that this source is a TeV emitter. Notably, such a hard $\gamma$-ray photon index safely rules out an extragalactic origin for 2FHL~J1745.1--3035, since no $\gamma$-ray detected extragalactic source has similar $\Gamma_\gamma$, and challenges our understanding of the emission mechanism that causes it.  

\begin{figure} 
 \centering 
\includegraphics[width=0.47\textwidth]{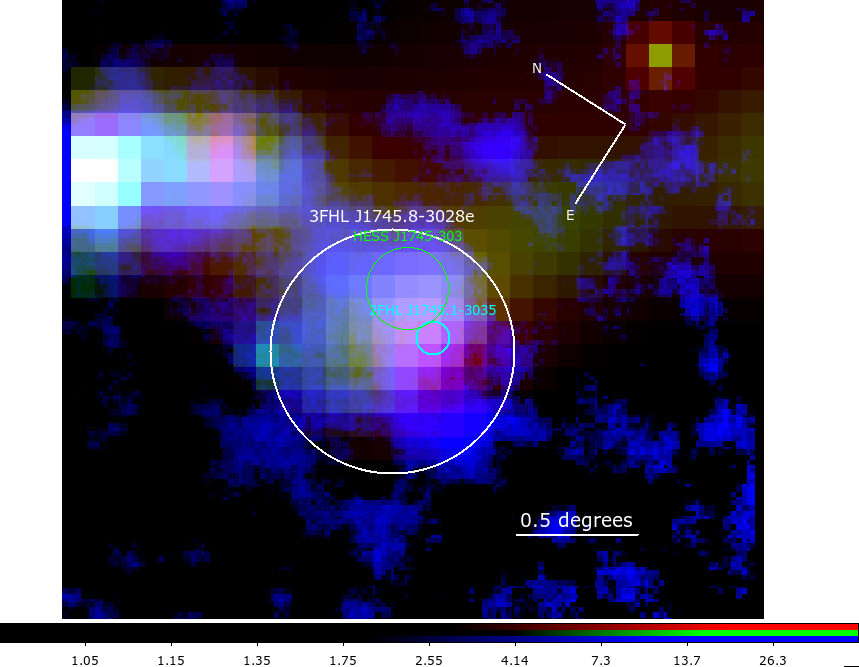} 
\caption{\normalsize 
Tricolor $\gamma$-ray image of the region where 2FHL~J1745.1--3035 is located. We plot in red the 2FHL counts ($>$50\,GeV), in green the 3FHL counts ($>$10\,GeV), and in blue the \hess\ flux in 0.03--100\,TeV band. We report in the figure the 2FHL, 3FHL, and \hess\ source locations: since the 3FHL (white circle) and \hess\ (green circle) sources are found to be extended, we plot a circle with a radius equal to the best-fit extensions reported in the catalogs. For the 2FHL source (cyan circle), which is found to be point-like, we use the 95\,\% confidence level radius.
}\label{fig:gamma_field}
\end{figure}

As shown in Figure~\ref{fig:gamma_field}, the 2FHL source is overlapping within the radius ($r$=0.53$^\circ$) of an extended object reported in two other \lat\ catalogs: namely, the Third Catalog of Hard Fermi-LAT Sources \citep[3FHL, which reports objects detected in the 10\,GeV--2\,TeV range;][source ID: 3FHL J1745.8-3028e]{ajello17} and the Fourth Fermi Large Area Telescope Catalog \citep[4FGL, which reports objects detected in the 50\,MeV--1\,TeV range;][source ID: 4FGL J1745.8-3028e]{abdollahi20}. The 3FHL source is associated with the 4FGL one. Both the 3FHL and the 4FGL source are modeled with a disk profile with radius $r$= 0.53 $\pm$ 0.02 $\pm$ 0.26$^\circ$ (where the first error is statistical and the second is systematic), as originally reported in the Fermi Galactic Extended Source Catalog \citep[FGES;][]{ackermann17}. There is however no source formally associated\footnote{The association procedure is based on a statistical Bayesian cross-correlation of a $\gamma$-ray catalog with other catalogs, as explained in detail in \citet{ackermann16}.} to  2FHL~J1745.1--3035 in the 4FGL and 3FGL catalogs. The 4FGL/3FHL \textit{Fermi} source has a TeV-detected counterpart of unknown origin, HESS J1745--303,  in the \hess\ Galactic plane survey \citep{hess18}. As shown in Figure~\ref{fig:fermi_SED}, the \hess\ source has a SED significantly different from the 2FHL one, a discrepancy that can be explained with the offset between the two sources, which may indicate different physical origins.

HESS J1745--303 has already been studied in several works in the literature. \citet{aharonian08} performed a multi-wavelength (including \xmm) study of the region where the \hess\ source is detected. They reported that no counterpart alone could fully explain the VHE emission detected by \hess: however, at least part of the TeV emission might be linked to either a supernova-remnant/molecular-cloud (SNR/MC) association or to a high-spin-down-flux pulsar, or a combination of the two. \citet{bamba09} reported the results of a \textit{Suzaku} observation of a part of the \hess\ region, but found no significant non-thermal X-ray emission. \citet{hui11} instead analyzed the combined \hess\ and 28--month \lat\ SED, but did not find a conclusive explanation for the VHE emission origin.

In summary, the complexity of the observed $\gamma$-ray field and the extreme properties of 2FHL~J1745.1--3035 make this source particularly interesting. Specifically, the $\gamma$-ray maps hint at a scenario where the emission observed and reported in the 2FHL catalog might not be entirely caused by the same object, or processes, causing the emission detected by \hess\ (and likely in the 4FGL catalog). To better characterize this intriguing 2FHL source, we therefore performed an extended, multi-X-ray observatory follow-up campaign.

\section{X-ray Data reduction and analysis}\label{sec:analysis}
We report in Table~\ref{tab:obs_summary} a summary of the five X-ray observations of 2FHL~J1745.1--3035 that have been performed between March 2001 and March 2023. In the following subsections, we describe in detail the data reduction and analysis for each instrument.

\begingroup
\renewcommand*{\arraystretch}{1.5}
\begin{table*}
\scalebox{1.0}{
\vspace{.1cm}
 \begin{tabular}{cccccc}
 \hline
 \hline
 Instrument & Sequence & Start Time  & Exposure & Net count rate  \\ 
& ObsID & (UTC) & ks & 10$^{-2}$ counts s$^{-1}$\\ 
  \hline    
 \xmm & 0103261301 & 2001-03-21T23:12:15 & 7.6; 7.6; 4.4 & 1.07$\pm$0.12; 1.29$\pm$0.13; 1.99$\pm$0.22 \\
 \xmm & 0782170601 & 2017-04-03T11:56:57 & 9.6; 9.6; 6.5 & 1.53$\pm$0.13; 1.28$\pm$0.12; 5.11$\pm$0.29 \\ 
 \xmm & 0886010401 & 2021-03-16T19:06:37 & 21.3; 21.3; 16.5 & 0.74$\pm$0.06; 0.96$\pm$0.07; 2.67$\pm$0.14 \\ 
 \cha & 23573 & 2022-05-20T08:52:56 & 22.8 & 0.59$\pm$0.05 \\
 \nus & 60301026002 & 2023-03-27T19:01:09 & 111 & 0.68$\pm$0.04; 0.74$\pm$0.05 \\ %
  \hline
	\hline
\end{tabular}}
	\caption{\normalsize Summary of the X-ray observations of the  2FHL~J1745.1--3035 counterpart.  The \xmm\ count rates are computed in the 0.6--10\,keV band, the \cha\ ones in the 0.6--8\,keV band, and the \nus\ ones are computed in the 3--70\,keV band. The \xmm\ exposures and count rates are the MOS1, MOS2 and pn ones, in this order; the \nus\ exposures and count rates are the FPMA and FPMB, in this order. All count rates are the net, background-subtracted ones.  Exposures are computed after removing high-background periods.
	}
\label{tab:obs_summary}
\end{table*}
\endgroup

\subsection{\xmm}\label{sec:xmm_data}
The source is in the field of view of three \xmm\ observations.

\begin{figure} 
 \centering 
 \includegraphics[width=0.46\textwidth]{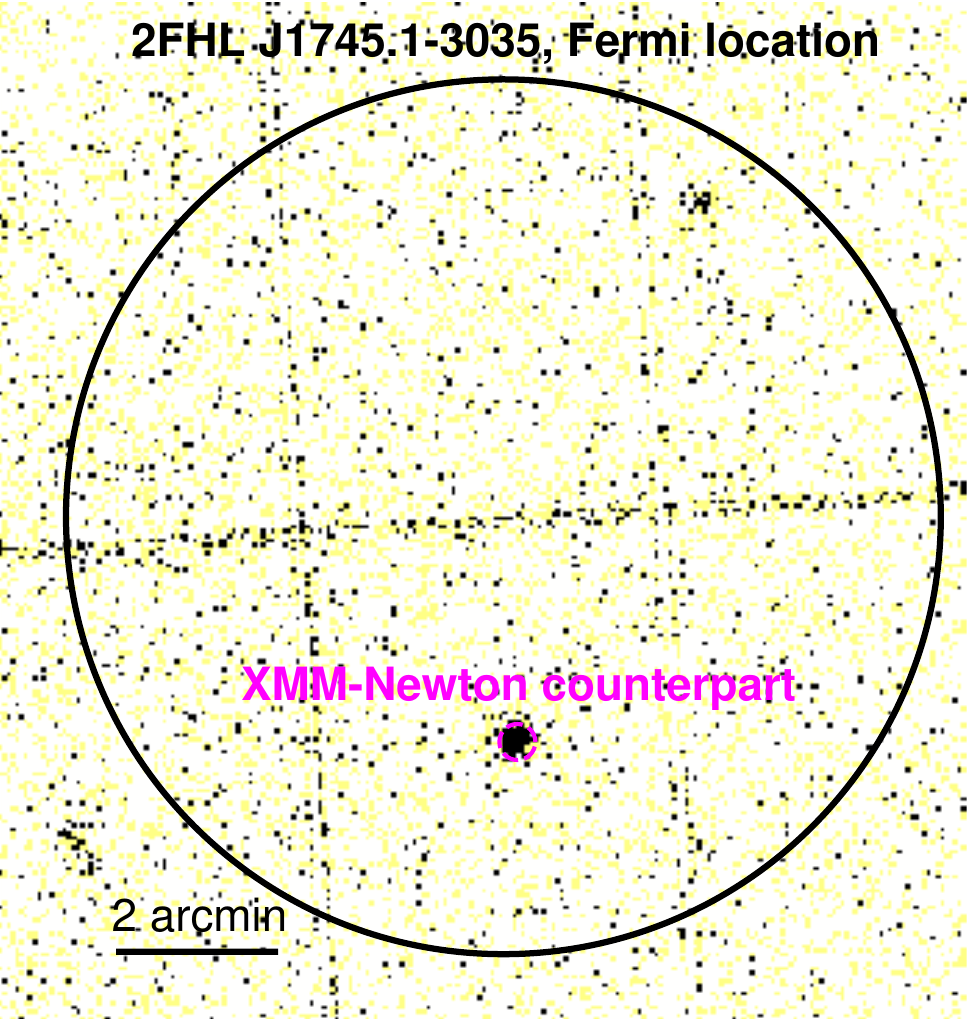} 
\caption{\normalsize 
XMM pn CCD 0.3--10\,keV image, centered on the 2FHL~J1745.1--3035 \lat\ position. The black circle shows the LAT 90\% confidence positional uncertainty ($r=5.4^\prime$). The X-ray counterpart is uniquely detected within the LAT uncertainty region.
}\label{fig:xmm_w_2FHL_region}
\end{figure}

\begin{enumerate}
    \item The first observation we studied is a 16\,ks one, taken in 2017 and aimed specifically at finding an X-ray counterpart for 2FHL~J1745.1--3035 (observation ID: 0782170601; PI: M. Ajello). In the \xmm\ cameras' field of view, the source was found at a distance from the 2FHL centroid (and therefore at an off-axis angle)  $\sim$3.5$^\prime$: the J2000 coordinates of the \xmm\ counterpart centroid are R.A. = 17:45:07.99 and Dec = --30:39:06.17. This source is reported in the fourth release of the \xmm\ source catalog\footnote{\url{http://xmmssc.irap.omp.eu/Catalogue/4XMM-DR13/4XMM_DR13.html}}, 4XMM-DR13 \citep{zolotukhin17,traulsen20,webb20}, with source ID 4XMM J174507.9-303906 (\url{http://xmm-catalog.irap.omp.eu/source/201032613010001}).   
    We report in Figure~\ref{fig:xmm_w_2FHL_region} an image of the \xmm\ pn 0.3--10\,keV observation with the \lat\ 90\,\% confidence positional uncertainty overlaid: it can be seen that the X-rays allow us to reliably identify a bright counterpart of the 2FHL source (magenta circle in Figure~\ref{fig:xmm_w_2FHL_region}). As a safety check, we nonetheless verified the properties of the faint source that can be noticed in the upper right part of the image. This a soft X-ray source with no emission above 2\,keV, which has a bright (G$_{\rm AB}$=12.5) counterpart in the \textit{Gaia} DR3 catalog \citep[\url{https://gea.esac.esa.int/archive/}; source ID: Gaia DR3 4056681582705849600][]{gaia16,gaia23}, where the source is flagged as a F-G class star (T$_{\rm eff}\sim$6000\,K) with 100\,\% probability. We thus safely rule it out as a possible counterpart of the 2FHL source.
    The detection of a bright, compact X-ray counterpart strongly supports a scenario where the 2FHL--detected emission does not have the same origin of the \hess\ one, which would explain why the two objects have such different SEDs. We also note that the following 
    \cha\ and \nus\ observations have been performed targeting the \xmm--detected object. Finally, we used the Aladin web tool\footnote{\url{aladin.cds.unistra.fr}} to search for possible counterparts of the X-ray source, but none was identified in the optical images down to magnitudes $R_{\rm AB}\sim$22.
    As reported in Table~\ref{tab:obs_summary}, this observation was affected by strong flares, therefore the observation MOS (pn) net exposure time is of $\sim$9.5\,ks ($\sim$6.5\,ks).
    \item The detection of an X-ray source in the \xmm\ 2017 image allowed us to perform a search in the \xmm\ archive aimed at finding other observations of the object. We find two such observations: the first one was taken in 2001, is 8\,ks long, and targeted the pulsar PSR B1742-30 (also known as PSR J1745-3040; observation ID: 0103261301; PI: F. Jansen). In this observation, our source is found at an off-axis angle $\sim$10$^\prime$.
    We performed a cross-match between the X-ray coordinates of our target and the Australia Telescope National Facility (ATNF) Pulsar Catalog \citep{manchester05}\footnote{The updated catalog is available at \url{http://www.atnf.csiro.au/research/pulsar/psrcat/}}. The closest source to our X-ray target is indeed PSR B1742-30. Based on the information in the ATNF Pulsar Catalog, PSR B1742-30 is at a $\sim$200\,pc distance from the Earth and has an estimated age of 546\,kyr. While a displacement of 10$^\prime$ (which is 0.6\,pc) could be reasonably expected as a consequence of a pulsar kick, a pulsar that old would be expected to be accompanied by a bow-shock nebula at the pulsar position in X-rays, while the relic nebula left behind would not be as compact as the source that we observe either in the X-rays nor in the 2FHL catalog. Furthermore, the pulsar E-dot as reported in the ATNF Pulsar Catalog is $\dot E$ = 8.5 $\times$ 10$^{33}$\,erg s$^{-1}$, a value significantly lower than those commonly measured in pulsars powering PWNe \citep[$\sim$10$^{36}$\,erg s$^{-1}$; see, e.g.,][]{mattana09,acero13}.
    We therefore conclude that the source we detect is likely not associated with the known pulsar PSR B1742-30, and would instead be a newly discovered pulsar.
    \item More recently, in 2021, our X-ray source was serendipitously observed during a 23\,ks observation of the Milky Way plane (observation ID: 0886010401; PI: G. Ponti). In this observation, the object we are studying is located at an off-axis angle $\sim$9$^\prime$.
\end{enumerate}

For all of the observations, the \xmm\ standard data reduction procedure has been performed using the SAS software, version 19.0.0. To extract the source spectra we used a circular region with radius $r_{\rm src}$=20$^{\prime\prime}$, while the background spectra were extracted from a circle having radius $r_{\rm bkg}$=45$^{\prime\prime}$ located in proximity of the source and visually inspected to avoid contamination from bright targets. Finally, given the different exposure times of the three observations, we adopted a different spectral binning for each of them: the 2001 and 2017 spectra are binned with 7 counts per bin, while in the 2021 observation we binned the MOS spectra with 10 counts per bin and the pn spectrum with 15 counts per bin.

\subsection{\cha}\label{sec:chandra_data}
2FHL~J1745.1--3035 was observed in May, 2022 with a $\sim$25\,ks \cha\ ACIS-I observation (PI: S. Marchesi). The dataset, obtained by the \cha\ X-ray Observatory, is contained in the \cha\ Data Collection (CDC) cdc.191~\dataset[doi:10.25574/cdc.191]{https://doi.org/10.25574/cdc.191}. We performed a standard data reduction using the CIAO software \citep{fruscione06}, version 4.14. We then followed the standard astrometric correction procedure reported in the \cha\ threads\footnote{\url{https://cxc.cfa.harvard.edu/ciao/threads/reproject_aspect/}} to correct the astrometry of our \cha\ image by using as a reference the USNO-A2.0 astrometric standards catalog \citep{monet98}. We find and apply offset corrections $\Delta$RA = RA$_{\rm Chandra}$ - RA$_{\rm USNO}$ = 0.471$^{\prime\prime}$ and $\Delta$Dec = Dec$_{\rm Chandra}$ - Dec$_{\rm USNO}$ = 0.278$^{\prime\prime}$.
We detect the \cha\ counterpart of 4XMM J174507.9-303906 at the position R.A. = 17:45:08.03 and Dec = --30:39:06.76, just 0.7$^{\prime\prime}$ away from the 4XMM source centroid.

\begin{figure*} 
 \centering 
 \includegraphics[width=0.99\textwidth]{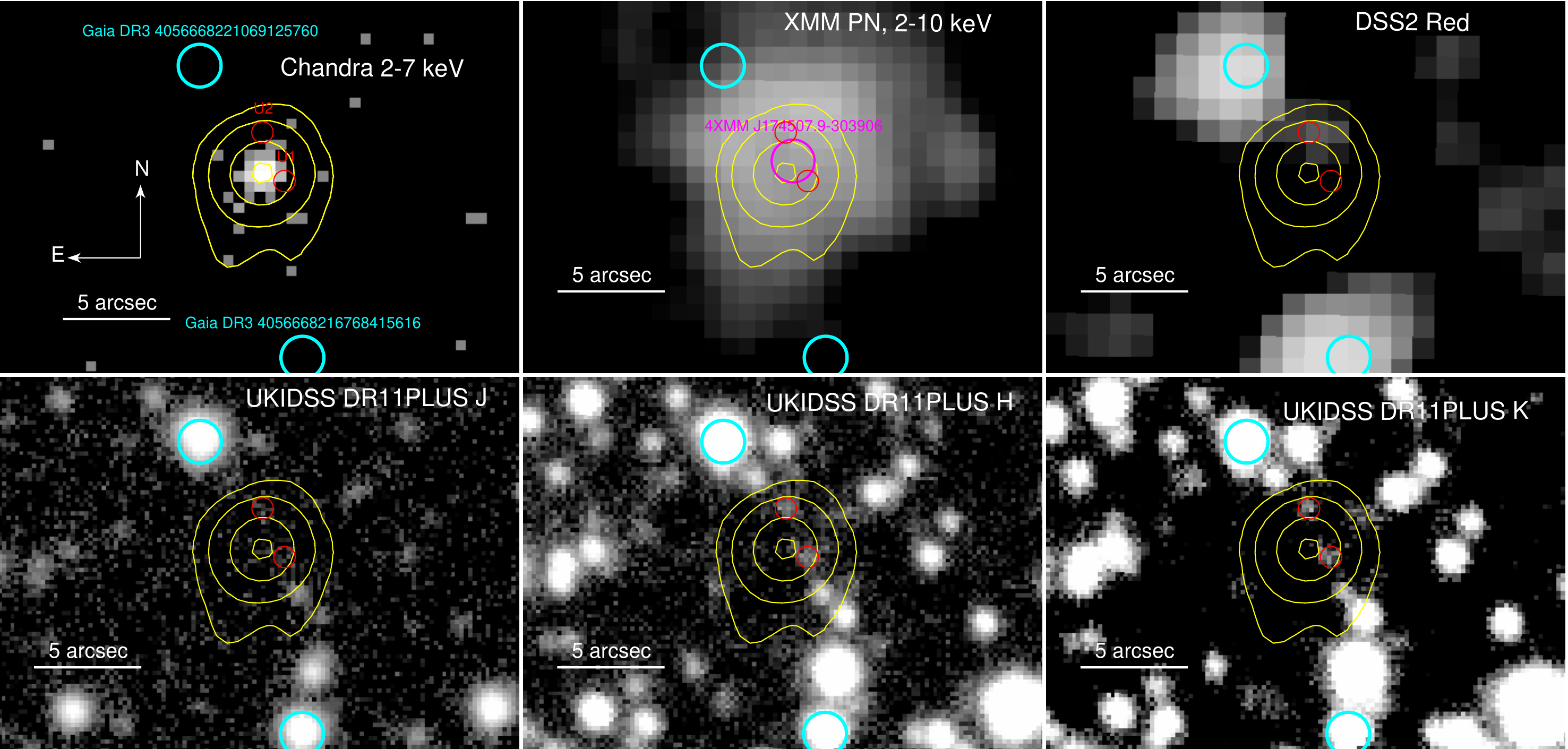} 
\caption{\normalsize Zoom-in of 4XMM J174507.9-303906 in the \cha\ 2--7\,keV image (top left), in the \xmm\ PN 2--10\,keV image (top center), in the DSS2-red survey \citep[top right;][]{mclean00} and in the UKIDSS survey \citep[bottom right;][]{skrutksie06} J, H, and K bands (bottom left, center, and right, respectively). In the \xmm\ image, we report the 4XMM catalog position of 4XMM J174507.9-303906.  For visual clarity, we overlay the smoothed \cha\ contours on the other images. The positions of two \textit{Gaia} sources and two UKIDSS sources are also shown as cyan and red circles, respectively.}\label{fig:X-ray_vs_optical_NIR}
\end{figure*}

We report in Figure~\ref{fig:X-ray_vs_optical_NIR} a zoom-in of 4XMM J174507.9-303906 in the \cha\ 2--7\,keV image (top left), in the smoothed \xmm\ PN 2--10\,keV image (top right), in the DSS2-red survey \citep[bottom left;][]{mclean00} and in the UKIRT Infrared Deep Sky Survey (UKIDSS) DR11PLUS J, H, and K band survey \citep[bottom left, center, and right;][]{warren07}. As it can be seen, no clear counterpart of the X-ray source is detected in the optical, where the two closest targets to the X-ray source are two objects both classified as stars with 100\,\% probability in the \textit{Gaia} DR3 catalog (these GAIA sources are plotted as cyan circles in the figure). There are instead two faint near-infrared sources in the proximity (distance $\lesssim$1$^{\prime\prime}$) of the \cha\ centroid, as reported in the UKIDSS--DR6 Galactic Plane Survey \citep{lucas08}\footnote{\url{https://vizier.cds.unistra.fr/viz-bin/VizieR-3?-source=II/316}}. We flag these sources as U1 (UKIDSS ID: J174507.94+429192822.5, H$_{\rm AB}$ = 18.1 $\pm$ 0.2; K$_{\rm AB}$ = 17.2 $\pm$ 0.2) and U2 (UKIDSS ID: J174508.02+429192824.8; K$_{\rm AB}$ = 17.2 $\pm$ 0.2) in Figure~\ref{fig:X-ray_vs_optical_NIR}. The faintness of these sources with respect to the X-ray brightness of our target, and the high source density of this field in the near infrared (as a reference, there are 28 UKIDSS sources within a 10$^{\prime\prime}$ radius from the \cha\ centroid) do not allow us to reliably associate either of the UKIDSS sources to the X-ray object.

We used the CIAO tool \texttt{specextract} to extract the \cha\ source spectrum from a circle having radius $r$=3$^{\prime\prime}$, while the background spectrum was extracted from a $r$=25$^{\prime\prime}$ circular region located nearby the source. The \cha\ spectrum was then binned with 10 counts per bin. In Section~\ref{sec:chandra_spatial} we will investigate in detail the presence of extended emission in the \cha\ image, and the subsequent implications on the nature of 2FHL~J1745.1--3035.

\subsection{\nus}\label{sec:nustar_data}
\begin{figure*} 
 \centering 
 \includegraphics[width=0.9\textwidth]{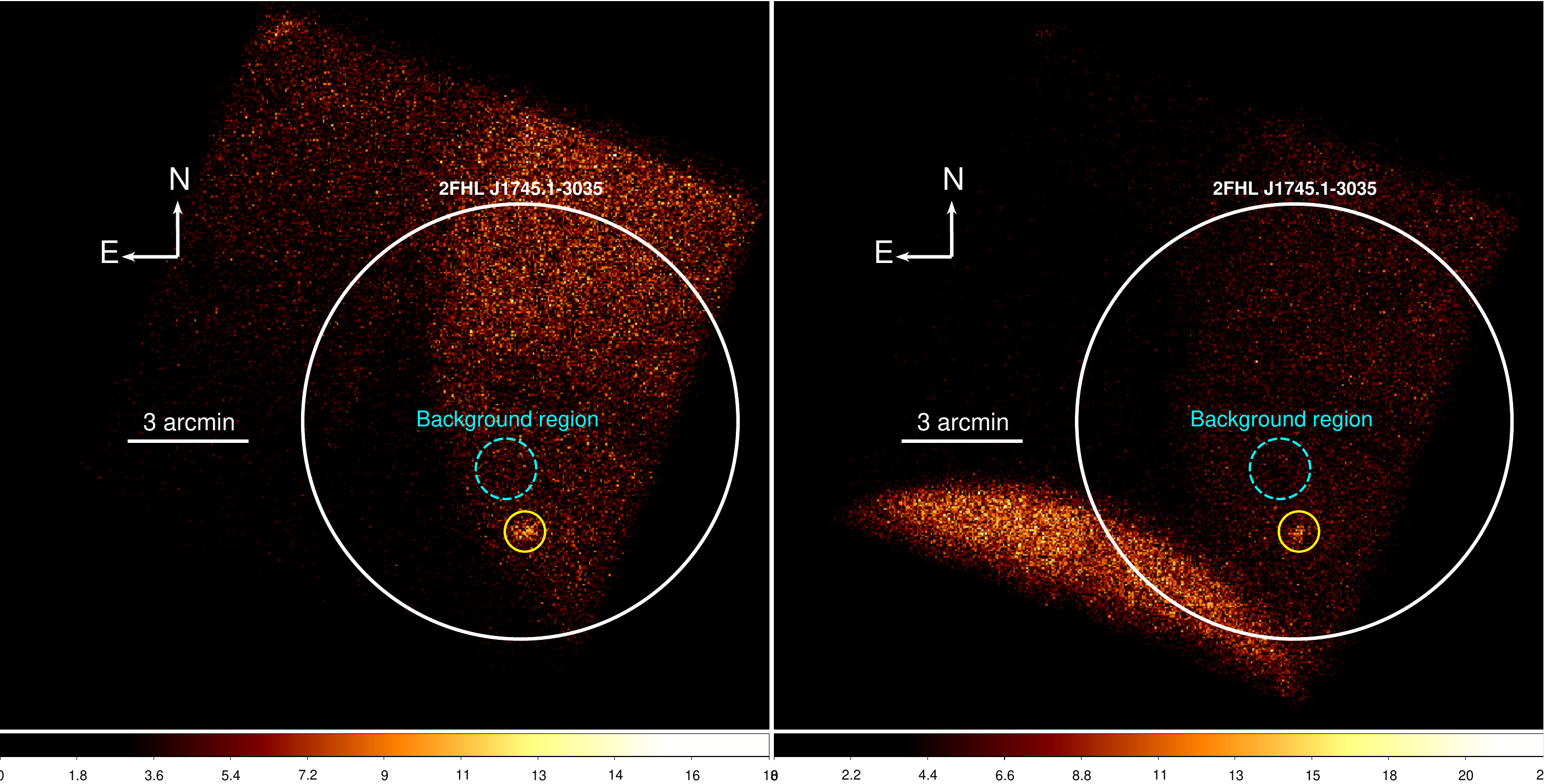} 
\caption{\normalsize 
\nus\ FPMA (left) and FPMB (right) observations of 2FHL~J1745.1--3035: the white circle marks the position of the \lat\ source, and has a radius corresponding to the $\gamma$-ray 90\,\% positional uncertainty ($r$=5.4$^\prime$). As it can be seen, stray light from the nearby Galactic Center is strongly affecting the observation: for this reason, the background region (dashed cyan circle) is selected as close as possible to the source (highlighted with a yellow circle), to ensure a background characterization as accurate as possible.
}\label{fig:nus_w_stray}
\end{figure*}

2FHL~J1745.1--3035 was observed by \nus\ in March, 2023. Our source is located in close proximity of the Galactic Center (longitude 358.481127$^\circ$, latitude -0.804527$^\circ$), which means that its \nus\ observations are strongly affected by stray light \citep{grefenstette21}. This is clearly visible in Figure~\ref{fig:nus_w_stray}, where we also report the region files which we used to extract the source and background spectra, with a radius of 30$^{\prime\prime}$ and 45$^{\prime\prime}$, respectively. Despite its challenging location, the source is clearly detected in both \nus\ cameras. Due to the high background, we binned the spectra with 75 counts per bin, to ensure an adequate signal-to-noise ratio.

\section{X-ray data analysis and interpretation }\label{sec:x-ray_analysis}
\subsection{Single-epoch X-ray fits and variability}\label{sec:single_epoch}
We perform single-epoch fits to search for potential variability in flux and photon index, both commonly observed in the X-ray spectra of PWNe \citep[e.g.,][]{pavlov01,klingler18,guest20}. In particular, X-ray variability in PWNe has already been observed over time-scales that vary from months to several years \citep[e.g., the PWN embedded in the SNR Kes 75;][]{ng08,livingstone11,reynolds18}, and it has been linked to physical processes in both the pulsar and the surrounding nebula. Such processes include synchrotron cooling of the PWN flow, resulting in a steepening of the X-ray photon index, or Kelvin--Helmoltz  instabilities in the shocked winds \citep{pavlov01}.

We note that we also searched for short-term variability within the observations, but we did not find any significant evidence of it. Interestingly, the 2017 and 2021 \xmm\ observations of  4XMM J174507.9-303906 are flagged as variable by the automatic pipeline used in the 4XMM catalog (\url{https://xcatdb.unistra.fr/4xmmdr13/xcatindex.html?detid=107821706010001} for the 2017 observation and \url{https://xcatdb.unistra.fr/4xmmdr13/xcatindex.html?detid=108860104010001} for the 2021 one). However, we performed a multi-band, multi-instrument analysis of the 2017 observation and found no evidence of simultaneous variability  in the three \xmm\ cameras (MOS1, MOS2 and PN) up to 6 keV. At energies greater than 6 keV, a flare is detected in all three cameras, but the same flare is still detected even at energies $>$10\,keV, where the contribution from the source becomes negligible and the emission is dominated by the background. We thus conclude that in the 2017 observation the variability reported in the 4XMM catalog can be ascribed to an episode of high background activity that commonly affects \xmm\ observations. As for the 2021 observation, we compared the MOS1, MOS2, and pn light curves in different bands and once again found no evidence of consistent cross-instrument variability. We report in the Appendix~\ref{app:light_curves} a summary of the light curves we produced for the 2017 and 2021 observations.

For each epoch of the X-ray observations, we fit the available data with an absorbed power law model, where the column density of the absorber was left free to vary. Since sources nearby the Galactic Center have been shown to be potentially affected by dust scattering \citep[see, e.g.,][]{jin18}, we performed a consistency check by fitting our data a first time using the \texttt{XSPEC} \texttt{XScat} model \citep{smith16}, and then a second time using the standard photoelectric absorption model \texttt{phabs}. Since the results are fully equivalent, we report below those obtained using the standard \texttt{phabs} model.

\begin{figure*} 
\begin{minipage}{0.31\textwidth} 
 \centering 
 \includegraphics[width=1\textwidth]{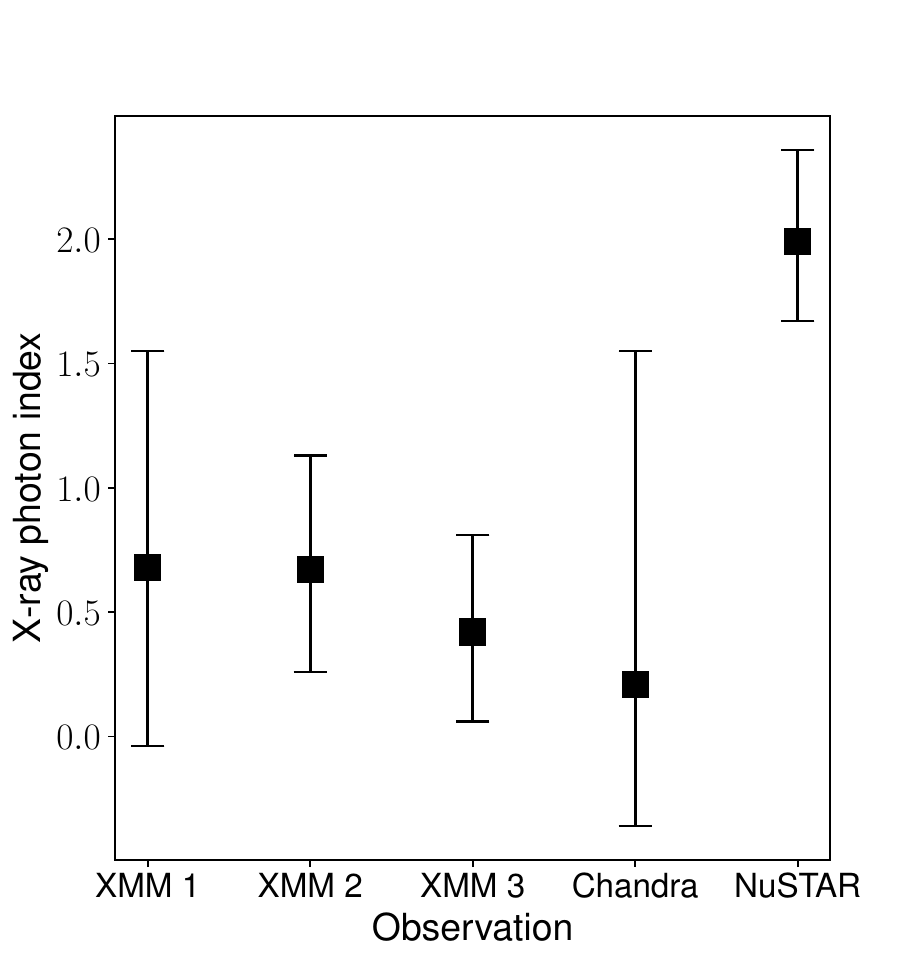} 
 \end{minipage} 
\begin{minipage}{0.35\textwidth} 
 \centering 
 \includegraphics[width=1\textwidth]{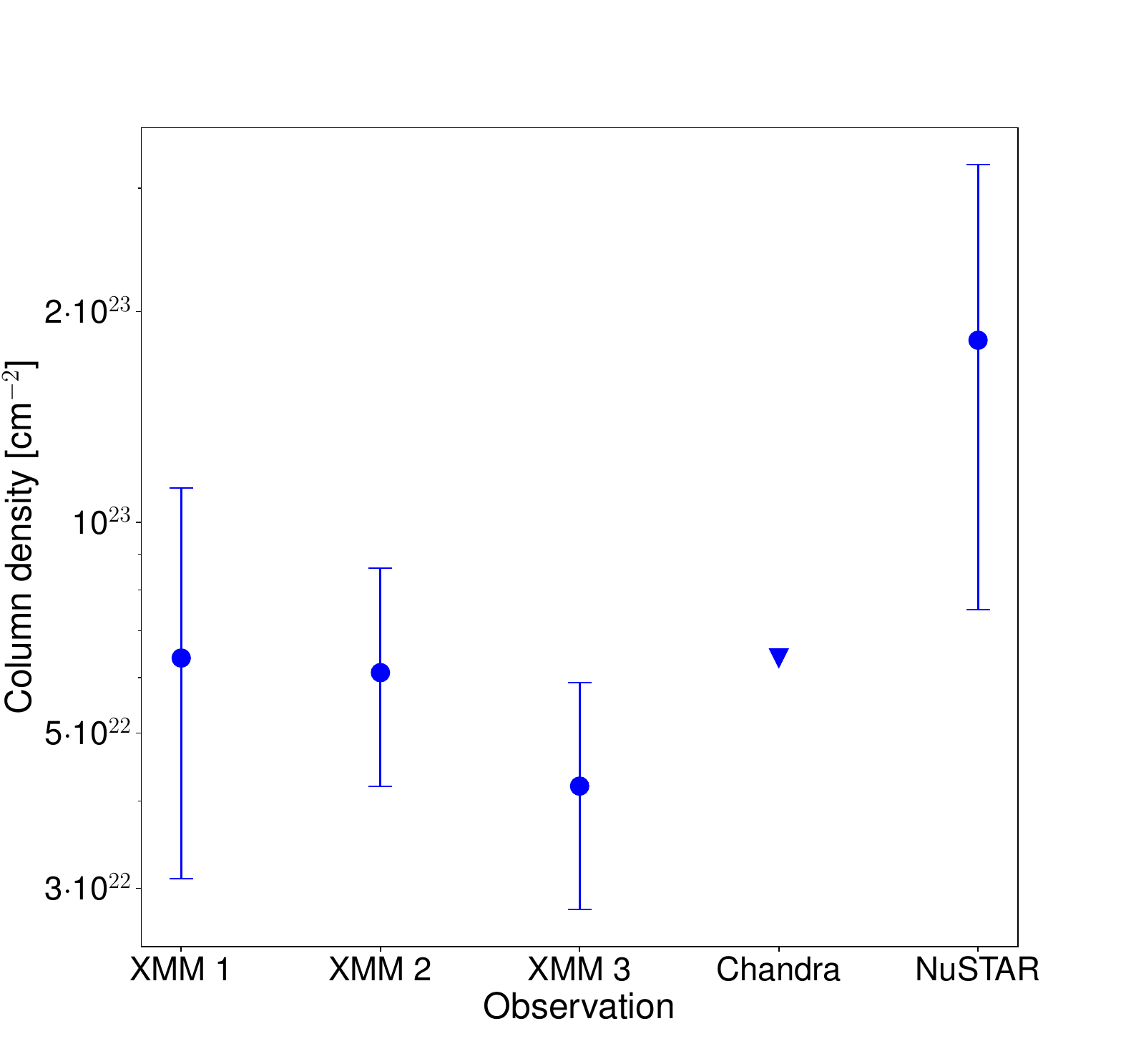}
 \end{minipage}
 \begin{minipage}{0.32\textwidth} 
 \centering 
 \includegraphics[width=1\textwidth]{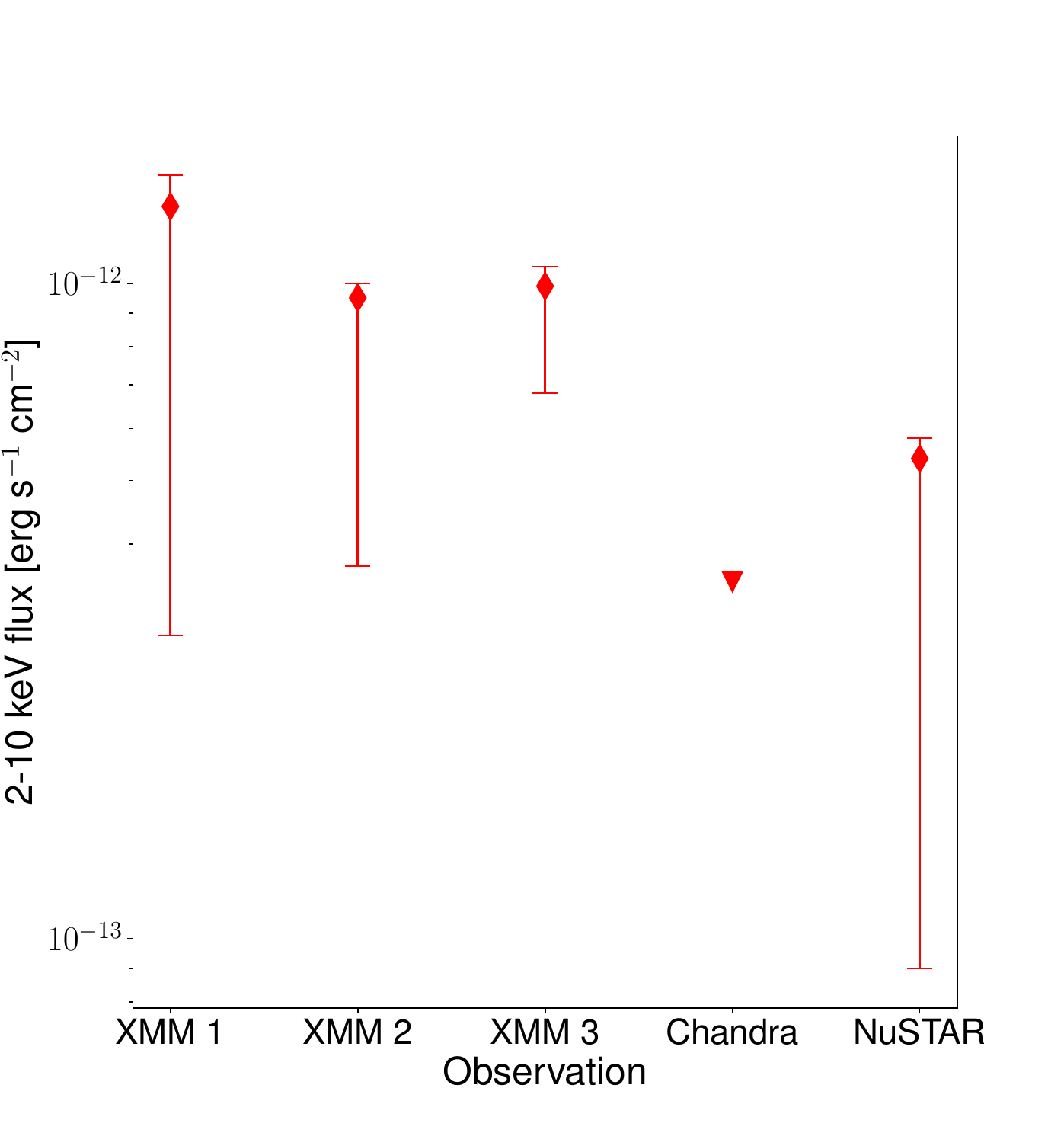}
 \end{minipage}
\caption{\normalsize 
Best-fit X-ray photon index (left), line-of-sight column density (center), and 2--10\,keV flux (right) for each of the five X-ray observations of 2FHL~J1745.1--3035. All observations are fitted with an absorbed single power law model. Downward triangles are used when only a 90\,\% confidence upper limit is measured.
}\label{fig:parameters_single_epoch}
\end{figure*}

When fitting \xmm\ and \nus\ observations, we also included cross-normalization constants to account for possible differences between different cameras. We also included in the model the absorption due to our own Galaxy, $N_{\rm H,Gal}$=9.7$\times$10$^{21}$\,cm$^{-2}$ \citep{kalberla05}. The best-fit results for each epoch are reported in Table~\ref{tab:single_epoch_fit}, while in Figure~\ref{fig:parameters_single_epoch} we plot the best-fit photon index, line-of-sight column density, and 2--10\,keV flux we measure in each epoch. The single-epoch spectra and the relative best-fit models are reported in Appendix~\ref{app:single_epoch}.

As it can be seen, the photon index and line-of-sight column density do not vary significantly in the three \xmm\ observations, as well as in the \cha\ one. In all cases, the X-ray photon index is significantly hard (varying in the range $\Gamma_{\rm X}$=[0.2--0.7]), and the X-ray emission is absorbed by material having column density \nhlos$\sim$4--6\,$\times$10$^{22}$\,cm$^{-2}$. The \nus\ best-fit results are instead different: most importantly, we measure an X-ray photon index that is significantly softer than those measured by the 0.5--10\,keV instruments, being $\Gamma_{\rm NuS}$=1.99$_{-0.32}^{+0.37}$. The fit to the \nus\ data also gives a slightly larger line-of-sight column density than the one measured with \cha\ and \xmm, which is N$_{\rm H,los,Nus}$=1.8$_{-1.1}^{+1.4}$\,$\times$10$^{23}$\,cm$^{-2}$. Such a discrepancy can however be explained with the fact that \nus\ lacks the $<$3\,keV coverage that is key to reliably constrain line-of-sight column densities smaller than 10$^{23}$\,cm$^{-2}$.

2FHL~J1745.1--3035 shows some marginal evidence for flux variability in the 2--10\,keV band. In particular, the 2022 \cha\ and the 2023 \nus\ data are $\sim$50\,\% fainter than the three \xmm\ observations taken between 2001 and 2021. However, the relatively large uncertainties on the flux measurements prevent us from studying this tentative trend in further detail. We also note that in the $\gamma$-rays no information on source flux variability is reported in the 2FHL catalog. There is also no observational evidence for flux variability in the 4FGL source associated with the \hess\ one and reported in the \lat\ data release 3 catalog \citep[DR3,][]{abdollahi22}, which is an incremental version of the 4FGL catalog published in 2020 and is based on the first 12 years of \lat\ observations between 50\,MeV and 1\,TeV. Specifically, the source variability index (computed as the sum of the log(likelihood) difference between the flux fitted in each time interval and the average flux over the full catalog interval) is 5.1. As a reference, a value greater than 18.48\footnote{Such a value corresponds to a 99\,\% confidence level in a $\chi^2$ distribution with seven degrees of freedom, see \citet{4fgl2020}} over 12 intervals indicates $<$1\,\% chance of being a steady source.

\begingroup
\renewcommand*{\arraystretch}{1.5}
\begin{table*}
\scalebox{1.0}{
\vspace{.1cm}
 \begin{tabular}{cccccc}
 \hline
 \hline
 Instrument & Observation date & $\chi^2$/d.o.f. & Photon Index & \nhlos & 2--10\,keV flux  \\ 
& (UTC) & & & 10$^{22}$ cm$^{-2}$ & 10$^{-12}$ erg s$^{-1}$ cm$^{-2}$ \\
  \hline    
 \xmm\ 1 & 2001-03-21 & 45.1/59 & 0.68$_{-0.72}^{+0.87}$ & 6.4$_{-3.3}^{+4.8}$    & 1.31$_{-1.02}^{+0.15}$ \\
 \xmm\ 2 & 2017-04-03 & 68.0/87 & 0.67$_{-0.41}^{+0.46}$ & 6.1$_{-1.9}^{+2.5}$    & 0.95$_{-0.58}^{+0.05}$ \\ 
 \xmm\ 3 & 2021-03-16 & 57.8/67 & 0.42$_{-0.36}^{+0.39}$ & 4.2$_{-1.4}^{+1.7}$    & 0.99$_{-0.31}^{+0.07}$ \\ 
 \cha    & 2022-05-20 &  4.7/10 & 0.21$_{-0.57}^{+1.34}$ & $<$6.4                 & $<$0.35\\
 \nus    & 2023-03-27 & 30.0/46 & 1.99$_{-0.32}^{+0.37}$ & 18.2$_{-10.7}^{+14.2}$ & 0.54$_{-0.45}^{+0.04}$\\ 
  \hline
	\hline
\end{tabular}}
	\caption{\normalsize Best-fit results of the single-epoch X-ray spectral fits of 2FHL~J1745.1--3035. All spectra are fitted with an absorbed single power law model, where \nhlos\ is the line-of-sight column density of the absorber. $\chi^2$/d.o.f. is the reduced $\chi^2$ of the fit, where d.o.f. are the degrees of freedom. In the \xmm\ (\nus) data the 2--10\,keV flux is computed from the pn (FPMA) dataset.
	}
\label{tab:single_epoch_fit}
\end{table*}
\endgroup

\subsection{Joint X-ray spectral fit}\label{sec:fit_joint}
Since we did not find any significant line-of-sight column density variability between the four soft X-ray observations, we jointly fit all X-ray spectra by tying this parameter among the different observations. Since we instead found evidence for potential X-ray flux variability, we added to the model a cross-normalization component to take this intrinsic variability into account. We then first fit the spectra with the same absorbed power law model we used in the single-epoch analysis. 

\begin{figure*} 
\begin{minipage}{0.49\textwidth} 
 \centering 
 \includegraphics[width=0.75\textwidth,angle=-90]{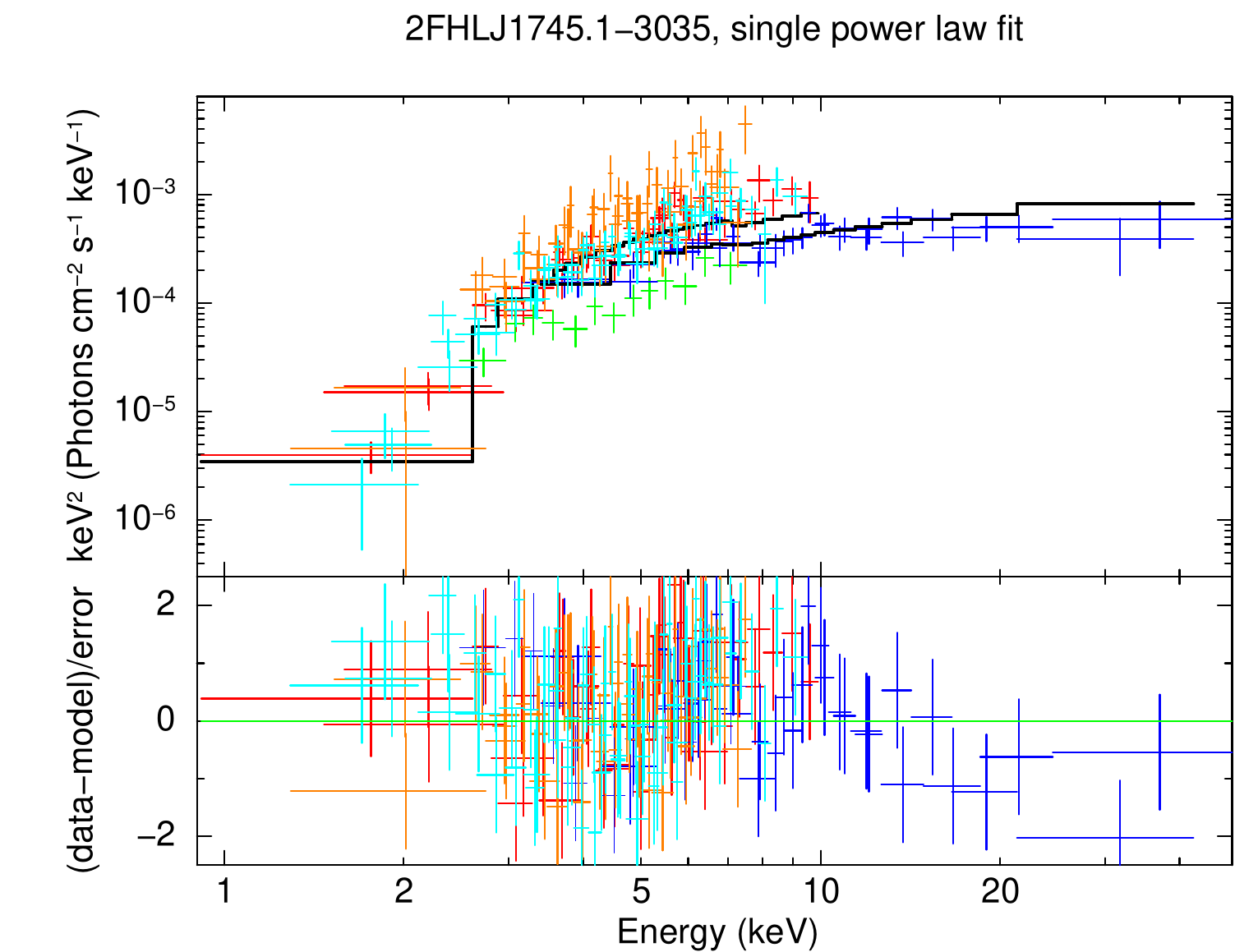} 
 \end{minipage} 
\begin{minipage}{0.49\textwidth} 
 \centering 
 \includegraphics[width=0.75\textwidth,angle=-90]{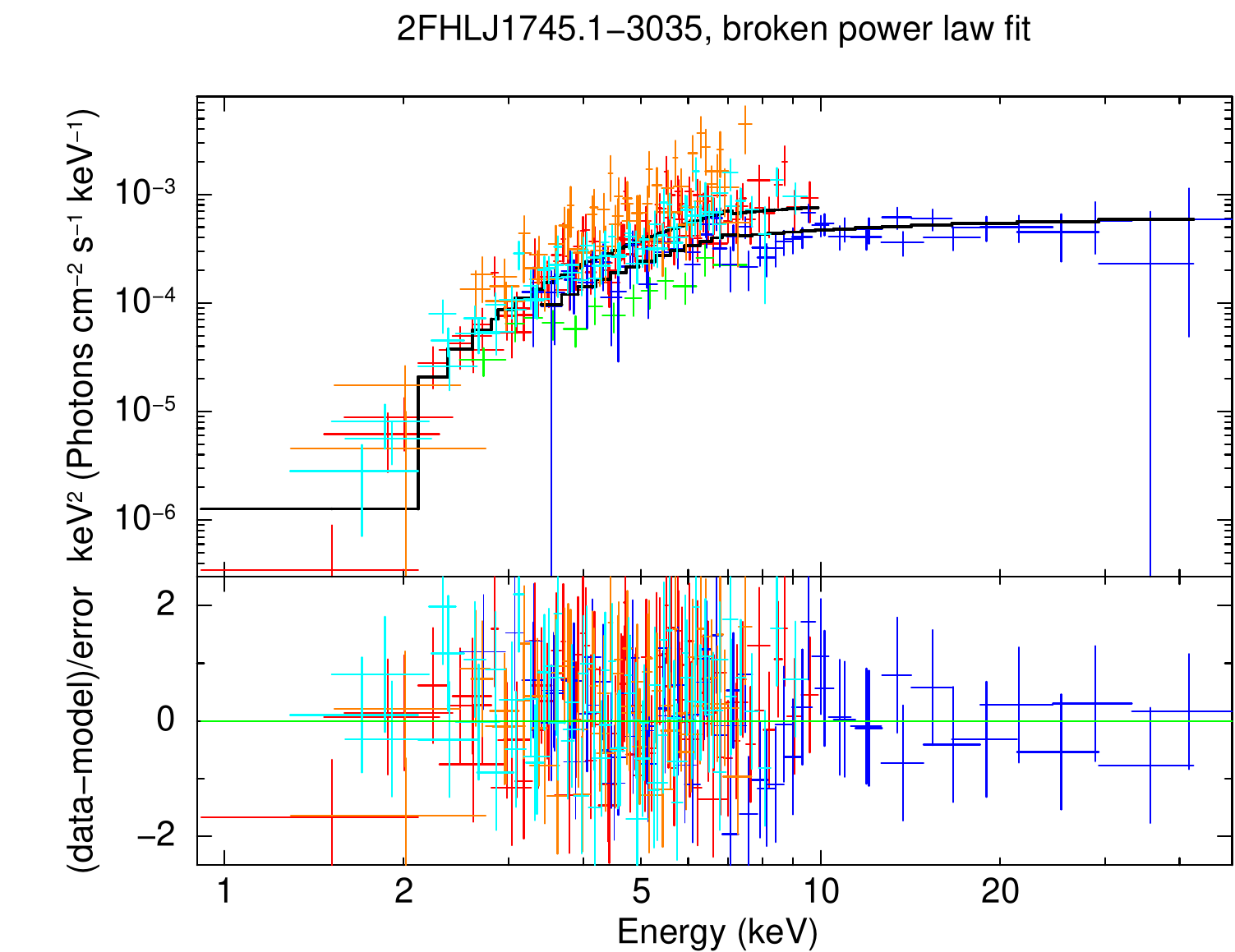}
 \end{minipage}
\caption{\normalsize 
Multi-epoch X-ray spectra of 2FHL~J1745.1--3035: the 2001, 2017, and 2021 \xmm\ observations are plotted in orange, red, and cyan, respectively; the 2022 \cha\ observation is plotted in green; the 2023 \nus\ observation is plotted in blue. The best-fit models (left: single power law; right: broken power law) are plotted as black solid lines. Residuals above 10 keV are clearly visible in the single power law best-fit model, while a significant improvement is visible in the broken power law fit. 
}\label{fig:joint_fit}
\end{figure*}

We measure a photon index $\Gamma_{\rm X}$=1.53$_{-0.15}^{+0.16}$ and a line-of-sight column density \nhlos=9.9$_{-1.1}^{+1.3}$\,$\times$10$^{22}$\,cm$^{-2}$. We report in Figure~\ref{fig:joint_fit}, left panel, the single power law best-fit model: significant residuals are visible in the ratio between the model and the data above 10\,keV. Furthermore, as expected, the best-fit photon index we measure lies between the values we obtain in the soft X-rays ($\Gamma_{\rm X}\sim$0.5, as shown in Table~\ref{tab:single_epoch_fit}) and the one measured using \nus, $\Gamma_{\rm X}$=1.99$_{-0.37}^{+0.32}$.

Given this observational evidence, we performed a new fit, this time using a broken power law model: notably, the presence of a break at $\sim$5--15\,keV in the X-ray spectra of PWNe has been reported in several works \citep[see, e.g.,][]{an14,nynka14,madsen15,madsen20,an19,bamba22}. 
Such a variation in $\Gamma_{\rm X}$ is an intriguing challenge to our understanding of PWN emission mechanisms, since PWNe SED models do not properly fit such an X-ray feature \citep[see, e.g.,][]{tanaka11}. Consequently, additional model components, such as for example different electron injection spectra or radially dependent PWN parameters would be required.

The parameterization of the broken-power law is 
\begin{equation}
   \frac{dN}{dE}=
\begin{cases}
   K E^{-\Gamma_{1}}& \text{if } E\leq E_{\rm b}\\
   K E_{\rm b}^{\Gamma_{2}-\Gamma_{1}}(E/1\,{\rm keV})^{-\Gamma_{2}})              & \text{otherwise}
\end{cases}
\end{equation}

Where $\Gamma_{1}$ and $\Gamma_{2}$  are the low- and high-energy photon indices, $K$ is the normalization parameter, and $E_{\rm b}$ is the energy of the break.

\begin{figure} 
 \centering 
 \includegraphics[width=0.99\linewidth]{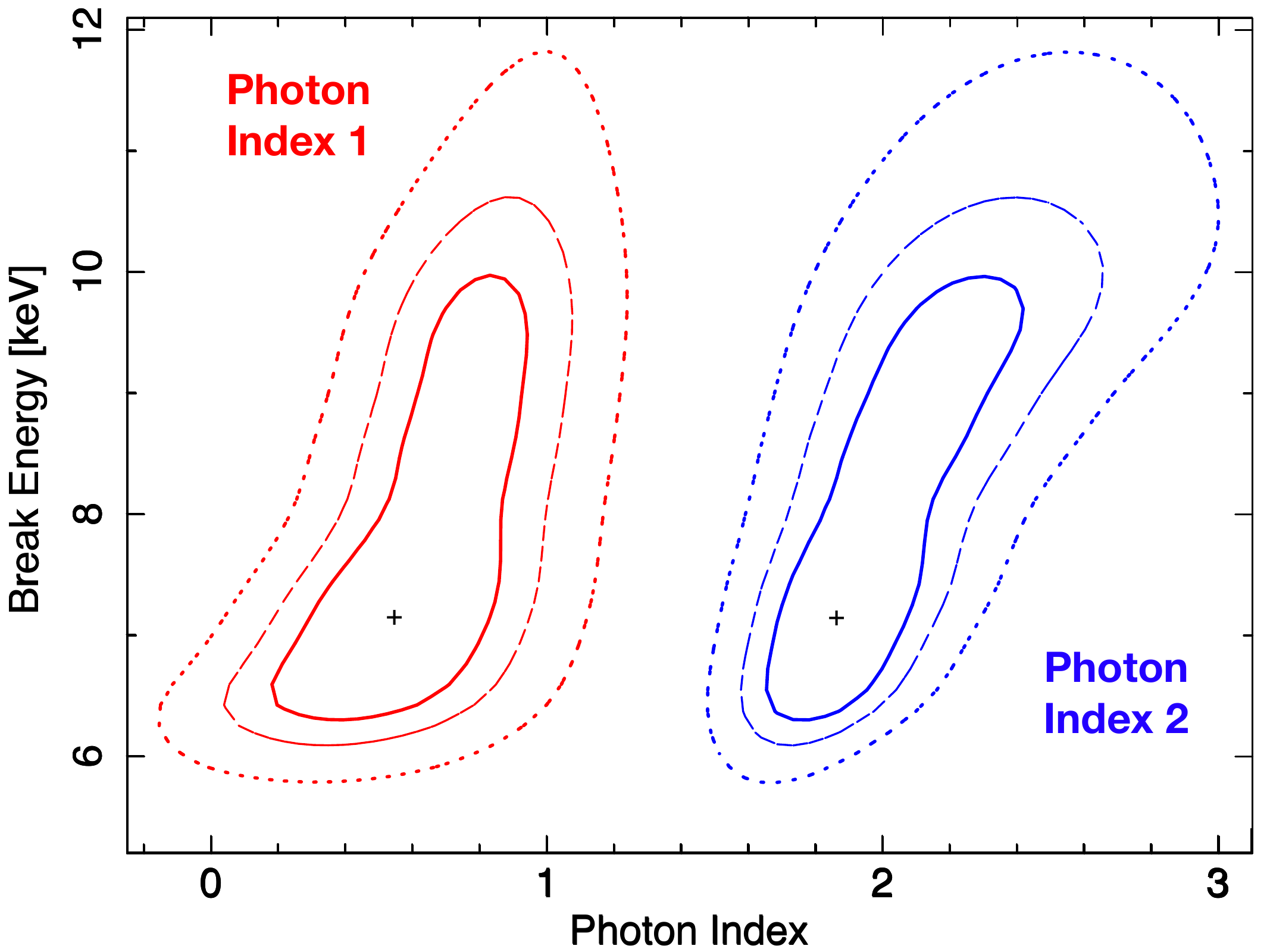} 
\caption{\normalsize 
68\,\% (solid line), 90\,\% (dashed line), and 99\,\% confidence levels for the broken power law energy break as a function of the photon index before the break, $\Gamma_{\rm 1}$ (red), and of the photon index after the break, $\Gamma_{\rm 2}$ (blue).
}\label{fig:contours_gamma_vs_Eb}
\end{figure}

The results of this fit are reported in Figure~\ref{fig:joint_fit}, right panel. As it can be seen, the data are much better constrained across the whole energy range, and in particular at energies E$>$10\,keV. We also measure a significant improvement in the fit, from $\chi^2$/d.o.f.=248.1/264 for the single power law model to $\chi^2$/d.o.f.=217.0/262 for the broken power law one. The photon index at energies lower than the break is $\Gamma_{\rm 1}$=0.56$_{-0.41}^{+0.42}$, in good agreement with the photon indices we measured in the \cha\ and \xmm\ spectra, while the second photon index is $\Gamma_{\rm 2}$=1.87$_{-0.24}^{+0.60}$, a value consistent with the one measured in the \nus\ spectra. Coherently, the best-fit break energy is E$_{\rm b}$=7.1$_{-0.9}^{+3.0}$\,keV, which is where the \nus\ contribution becomes the dominant one. In Figure~\ref{fig:contours_gamma_vs_Eb} we report the confidence contours of the break energy as a function of both photon indices: $\Gamma_{\rm 1}$ and $\Gamma_{\rm 2}$ are different at the $>$99\,\% confidence level, a result that further favors the broken power law scenario over the single power law one. We also note that the best-fit column density obtained with the broken power law fit, \nhlos=5.1$_{-1.4}^{+1.4}$ $\times$ 10$^{22}$\,cm$^{-2}$, is significantly larger than the one expected on the line of sight from the neutral Hydrogen measurements by \citet[][N$_{\rm H,Gal}$=9.7$\times$10$^{21}$\,cm$^{-2}$]{kalberla05}. Such a result supports a scenario where the X-ray emission is taking place in a dense, gas-rich environment.

Finally, as a consistency check, and in particular to partially rule out an SNR origin for 2FHL~J1745.1--3035, we fitted our data with the \texttt{nei} non-equilibrium ionization collisional plasma model, which is commonly used when modeling the X-ray spectra of SNRs \citep[see, e.g.,][]{harrus01,lazendic06,katsuda09,prinz13,lehay20,eagle22b}. We report the best-fit results in Table~\ref{tab:multi_epoch_fit}: while the best-fit statistic ($\chi^2$/d.o.f.=225.3/262) is significantly better than the one obtained using the single power law model, and fairly consistent with the one we obtain using the broken power law model, the best-fit temperature ($kT$=34.4$_{-9.1}^{+24.4}$ keV) is fully unphysical. As a reference, \citet{lehay20} analyzed the X-ray spectra of 43 SNR of different ages, explosion energies, and circumstellar medium densities: none of them was found having $kT\geq$4\,keV. Such a result reliably rules out a thermal SNR scenario for 2FHL~J1745.1--3035. 

\begingroup
\renewcommand*{\arraystretch}{1.5}
\begin{table*}
\scalebox{0.94}{
\vspace{.1cm}
 \begin{tabular}{ccccccccc}
 \hline
 \hline
 Model & $\chi^2$/d.o.f. & \nhlos & $\Gamma_{\rm 1}$ & $\Gamma_{\rm 2}$ & E$_{\rm b}$ & $kT$ & $\tau$ & $Z$\\ 
& & 10$^{22}$ cm$^{-2}$ & & & keV & keV & 10$^{10}$ s cm$^{-3}$ & $Z_\odot$ \\
  \hline    
Single Power Law & 248.1/264 & 9.0$_{-1.1}^{+1.3}$ & 1.53$_{-0.15}^{+0.16}$ & -- & -- & -- & -- & -- \\
Broken Power Law & 217.0/262 & 5.1$_{-1.4}^{+1.4}$ & 0.56$_{-0.41}^{+0.42}$ & 1.87$_{-0.24}^{+0.60}$ & 7.1$_{-0.9}^{+3.0}$ & -- & -- & -- \\
\texttt{nei} & 225.3/262 & 8.9$_{-1.4}^{+0.9}$ & -- & -- & -- & 34.4$_{-9.1}^{+24.4}$ & $<$5.95 & 0.24$_{-0.11}^{+0.11}$\\
  \hline
	\hline
\end{tabular}}
	\caption{\normalsize Best-fit results of the multi-epoch X-ray spectral fits of 2FHL~J1745.1--3035, for the three different models discussed in the text. $\chi^2$/d.o.f. is the reduced $\chi^2$ of the fit, where d.o.f. are the degrees of freedom; \nhlos\ is the line-of-sight column density of the absorber; $\Gamma_{\rm 1}$ is the photon index of the single power law model, and the photon index before the break of the broken power law model; E$_{\rm b}$ and $\Gamma_{\rm 2}$ are the energy break and the photon index at energies larger than it in the broken power law model, respectively; finally, $kT$, $\tau$, and $Z$ are the best-fit temperature, ionization time scale, and metallicity, derived using the \texttt{nei} thermal model.
	}
\label{tab:multi_epoch_fit}
\end{table*}
\endgroup

\subsection{Chandra spatially resolved analysis}\label{sec:chandra_spatial}
\begin{figure*} 
\begin{minipage}{0.49\textwidth} 
 \centering 
 \includegraphics[width=0.99\textwidth]{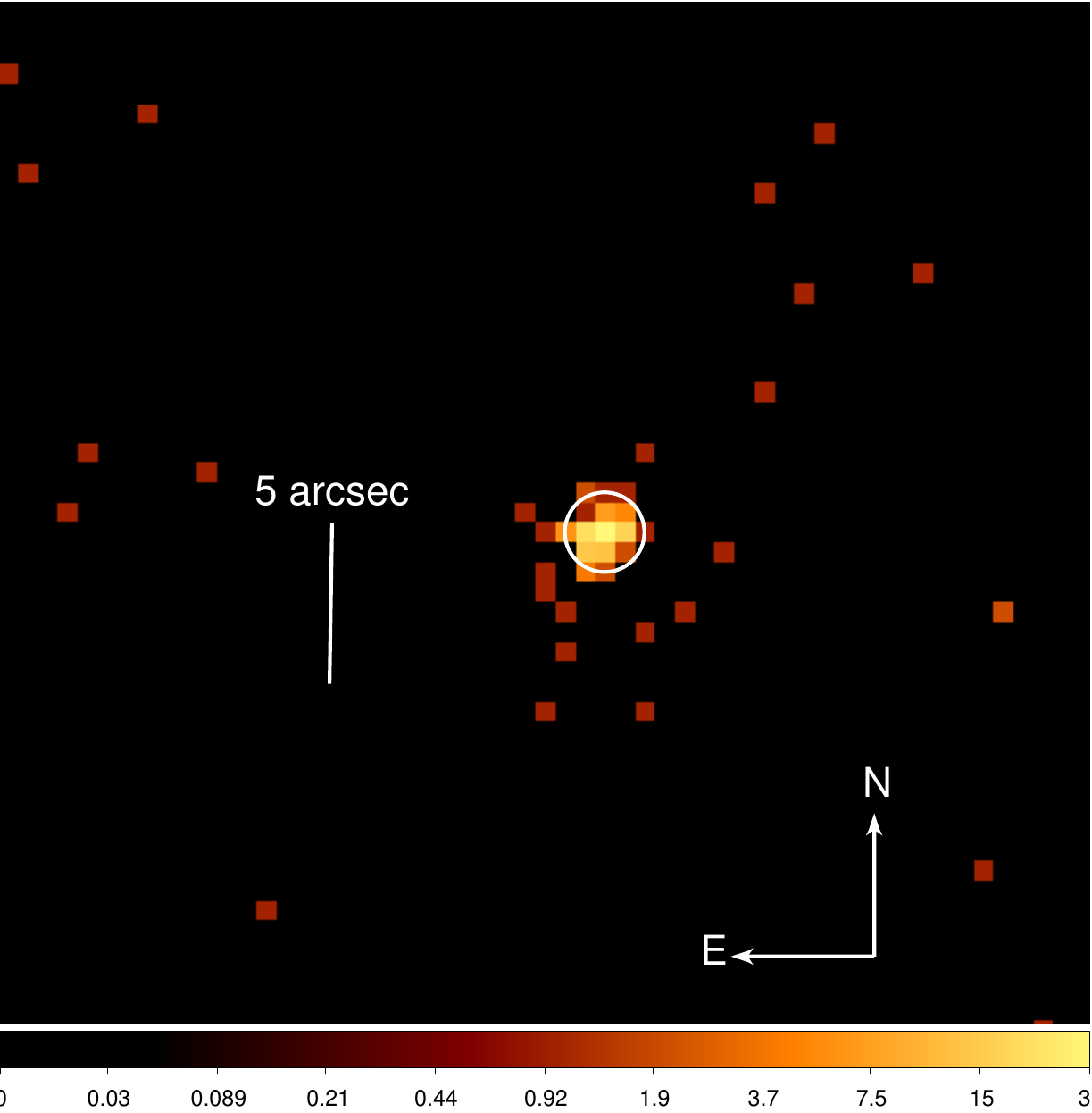} 
 \end{minipage} 
\begin{minipage}{0.49\textwidth} 
 \centering 
 \includegraphics[width=0.99\textwidth]{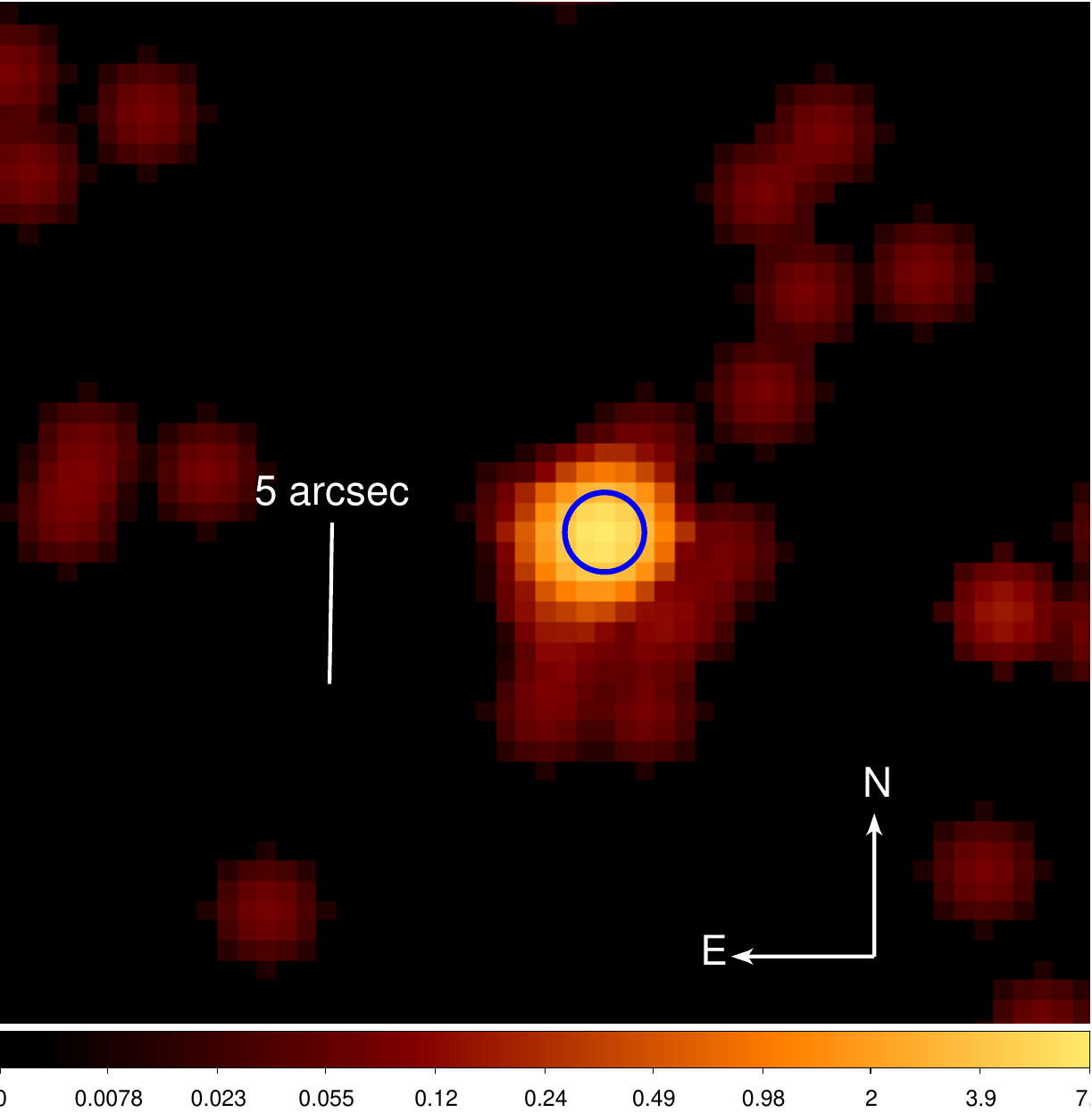}
 \end{minipage}
 \begin{minipage}{0.49\textwidth} 
 \centering 
 \includegraphics[width=0.99\textwidth]{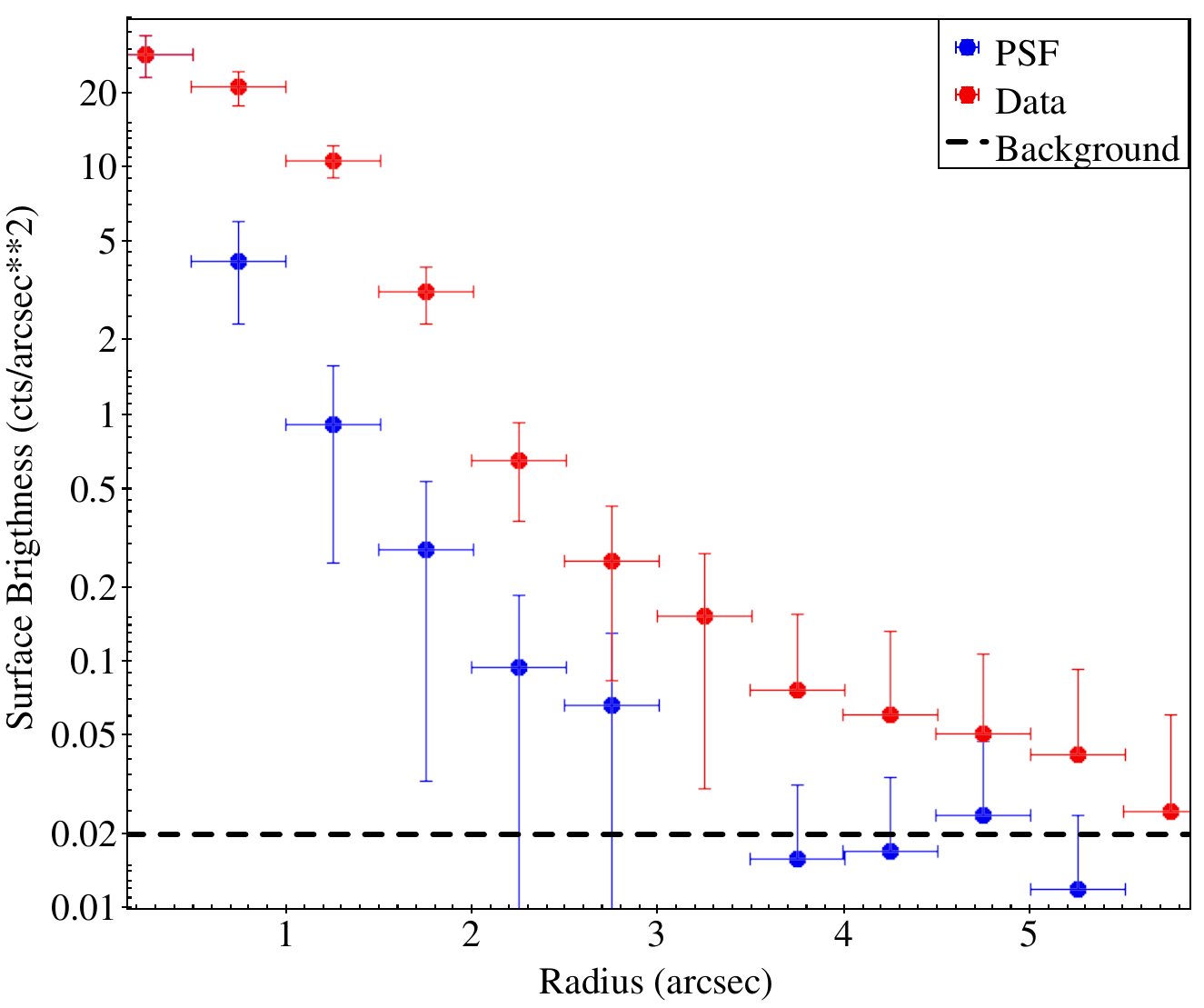}
 \end{minipage}
 \begin{minipage}{0.49\textwidth} 
 \centering 
 \includegraphics[width=0.99\textwidth]{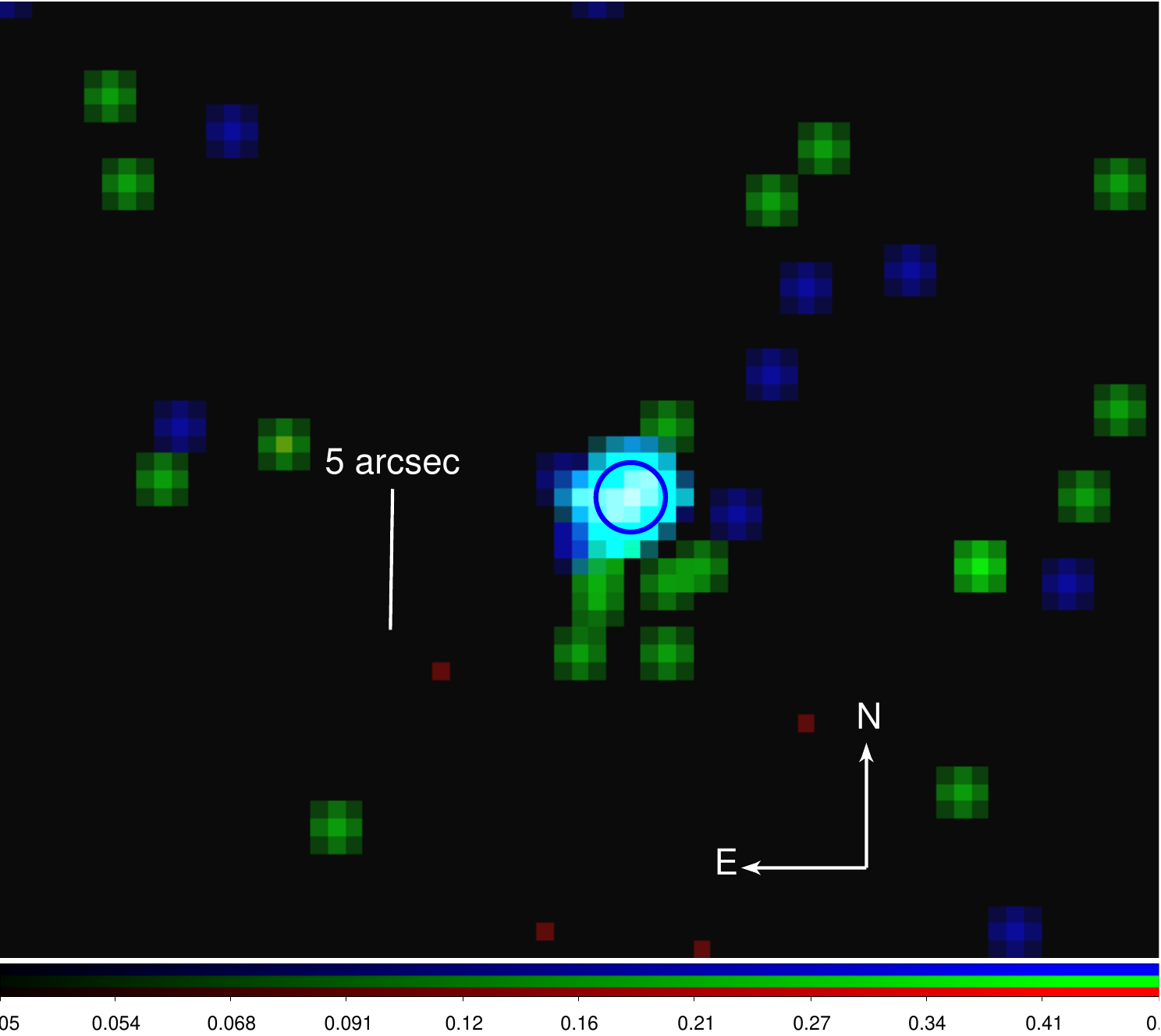}
 \end{minipage}
\caption{\normalsize 
Top panels: unbinned (left) and smoothed (center) \cha\ ACIS-I 2--7\,keV image of 2FHL~J1745.1--3035. In both images, we plot a circle centered on the brightest pixel and having a radius corresponding at the \cha\ ACIS on-axis 90\,\% encircled energy fraction (0.98$^{\prime\prime}$). Bottom panels: on the left we show the surface brightness of our target (red) as a function of the radius: for comparison we also show the expected trend for the \cha\ PSF (blue), normalized so that in the first bin the PSF surface brightness is equal to the measured one, as well as the average background surface brightness (dashed black line); the error bars represent the Poissonian error of the counts inside each annulus. On the right, we report a smoothed tricolor (red: 0.5--2\,keV; green: 2--4.5\,keV; blue: 4.5--7\,keV) image of 2FHL~J1745.1--3035. The fainter, diffuse emission is found to be softer than the central, unresolved one.
}\label{fig:chandra_extended}
\end{figure*}

The X-ray counterpart of 2FHL~J1745.1--3035 does not show clear evidence for extended emission in the \nus\ image or in the \xmm\ images, which implies that the X- and $\gamma$-ray emitter must be compact. Extended PWN emission is however commonly observed with \cha, and indeed a first visual evidence is visible in our \cha\ ACIS-I observation as well. As shown in the top panels of Figure~\ref{fig:chandra_extended}, both the unbinned \cha\ image and the smoothed one present evidence for asymmetric extended emission, with a faint tail of emission detected Southwards of the source center.

To quantitatively search for extended emission in 2FHL~J1745.1--3035, we perform a comparison between the 2--7\,keV surface brightness radial distribution and the one obtained from a simulated PSF in the same band, computed using the \texttt{ChaRT} and \texttt{MARX 5.5.1} tools \citep[see e.g.,][for a detailed description of this same approach in nearby active galactic nuclei]{fabbiano17,jones20,ma20,traina21,silver22}. We report in Figure \ref{fig:chandra_extended}, bottom left panel, the 2--7\,keV surface brightness radial profile of 2FHL~J1745.1--3035, which we computed from a set of 11 annuli, where the two radii increase by 0.5$^{\prime\prime}$ in each consecutive annulus, and the maximum external radius is 6.5$^{\prime\prime}$, and the corresponding expected PSF profile which one would observe for a point-like source. As it can be seen, this quantitative analysis confirms the first qualitative impression, since the source surface brightness exceeds the one predicted for a point-like scenario up to $\sim$5$^{\prime\prime}$. As a caveat, we note that it has been shown \citep{primini11}, that that the ray tracing simulations generated by MARX will have a roughly correct total intensity but tend to have PSF wings that are broader than observations and a PSF core narrower than observed, an effect that could in principle affect our simulation as well.

Finally, the low count statistic of the \cha\ spectrum prevents us from performing an energy dependent surface brightness profile, as well as a spatially resolved spectral fit. Nonetheless, a simple tricolor image such as the one we report in Figure~\ref{fig:chandra_extended}, bottom right panel, gives a first indication of the presence of two distinct sources of emission: a harder, unresolved one at the center (likely the pulsar) and a fainter, softer diffuse emission in the outskirts (in our scenario, the pulsar wind nebula). We note that \citet{li08} analyzed the \cha\ and \xmm\ spectra of a sample of pulsars and PWNe, and found tentative evidence for both direct correlation between the X-ray photon index of the pulsar and its age (i.e., younger pulsars have harder photon indices) and anti-correlation between the PWN photon index and the pulsar age (i.e., younger PWNe have softer photon indices). Both these trends support a young PWN scenario for 2FHL~J1745.1--3035, based on the tentative observational evidence we find in the \cha\ image.

\section{Spectral energy distribution modeling and physical characterization}\label{sec:SED_fit}

In order to investigate the origin of the observed high-energy emission, we use the NAIMA Python package \citep[v0.10.0,][]{naima}, which computes the radiation from a single non-thermal relativistic particle population. 
For the particle distribution in energy, we assume a power law shape with an exponential cut-off,
\begin{equation}
f(E) = A \bigg(\frac{E}{E_0}\bigg)^{-\alpha} \exp\big({-{\frac{E}{E_{c}}}}\big) \\
\end{equation}
We then test a combination of free parameters (namely the normalization $A$, index $\alpha$, and energy cut-off $E_{c}$) that can best explain the broadband spectrum. We consider three photon fields in all Inverse Compton Scattering calculations in this section, the Cosmic Microwave Background (CMB), a near infrared (NIR) and a far infrared (FIR) stellar photon field, set to the \texttt{GALPROP} values measured from a Galactic radius of 6.5\,kpc.

Due to the complex $\gamma$-ray emission detected in this region, it is plausible that there is source confusion from more than one object generating the observed $\gamma$-rays (as discussed in Section~\ref{sec:2FHLJ1745} and visually shown in Figure~\ref{fig:fermi_SED}). Indeed a reasonable broadband representation for the X- and $\gamma$-ray data requires more than one particle population and the predicted properties depend on the dominant $\gamma$-ray spectral properties we assume. We are unable to determine the presence or level of contribution from low- and high-energy components using the $\gamma$-ray data alone, so we explore two combinations of data for modeling: 1) radio, X-ray, and $\gamma$-ray data assuming the extended 4FGL and \hess\ sources are the dominant $\gamma$-ray counterparts and 2) combine the same radio and X-ray data to the point-like 2FHL data. We assume two scenarios of particle distributions in each case: two leptonic populations or a lepto-hadronic population. In an effort to constrain radio synchrotron contributions, we estimate an upper limit on the radio flux from the 1.28\,GHz MEERKAT Galactic Center radio observation\footnote{\url{https://archive-gw-1.kat.ac.za/public/repository/10.48479/fyst-hj47/index.html}, \citep{meerkat22}.} \citep{heywood2022}. We discuss the results of the broadband modeling in the following sections. The model parameters that can best reproduce the two data combinations are provided in Table~\ref{tab:naima}.

\subsection{Leptonic Scenario}
In the case of two-leptonic particle distributions, the radio--X-ray--2FHL data combinations is reasonably reproduced (as shown in the left panel of Figure~\ref{fig:naimamodels_2fhl}), while no combination of parameters can adequately describe the SED composed by the radio, X-ray, 4FGL and \hess\ data, and in particular the 4FGL data (Figure~\ref{fig:naimamodels}, left panel). This is due to the very low radio flux upper limit that forces the electron index to be harder than what is inferred by the 4FGL data ($\sim 2$). Introducing another fitting parameter such as the index to the energy cut-off (i.e., an index after the break deviating from the assumed value of 1 here) may be able to reconcile the poor fitting in the Fermi band. An index after the cutoff that is smaller than one, for instance, would soften the cut-off shape in the synchrotron peak for the low-energy population (Population~1 in left panel of Figure~\ref{fig:naimamodels}). A final possibility would be that the particle population is more complicated than what is explored here. 

As mentioned above, the radio, X-ray and 2FHL $\gamma$-ray data combination is instead reasonably reproduced by the two-leptonic model (left panel of Figure~\ref{fig:naimamodels_2fhl}). Assuming the two-leptonic scenario for both data combinations, the predicted properties are consistent with an energetic, young and/or more evolved PWN where the extended 4FGL and \hess\ $\gamma$-ray data may be an extended relic PWN and the point-like 2FHL emission may be a compact, young PWN located close to the pulsar.

\subsection{Lepto-Hadronic Scenario}
\begin{figure*}[ht!]
\begin{minipage}{0.49\textwidth} 
\centering
\includegraphics[width=1.0\linewidth]{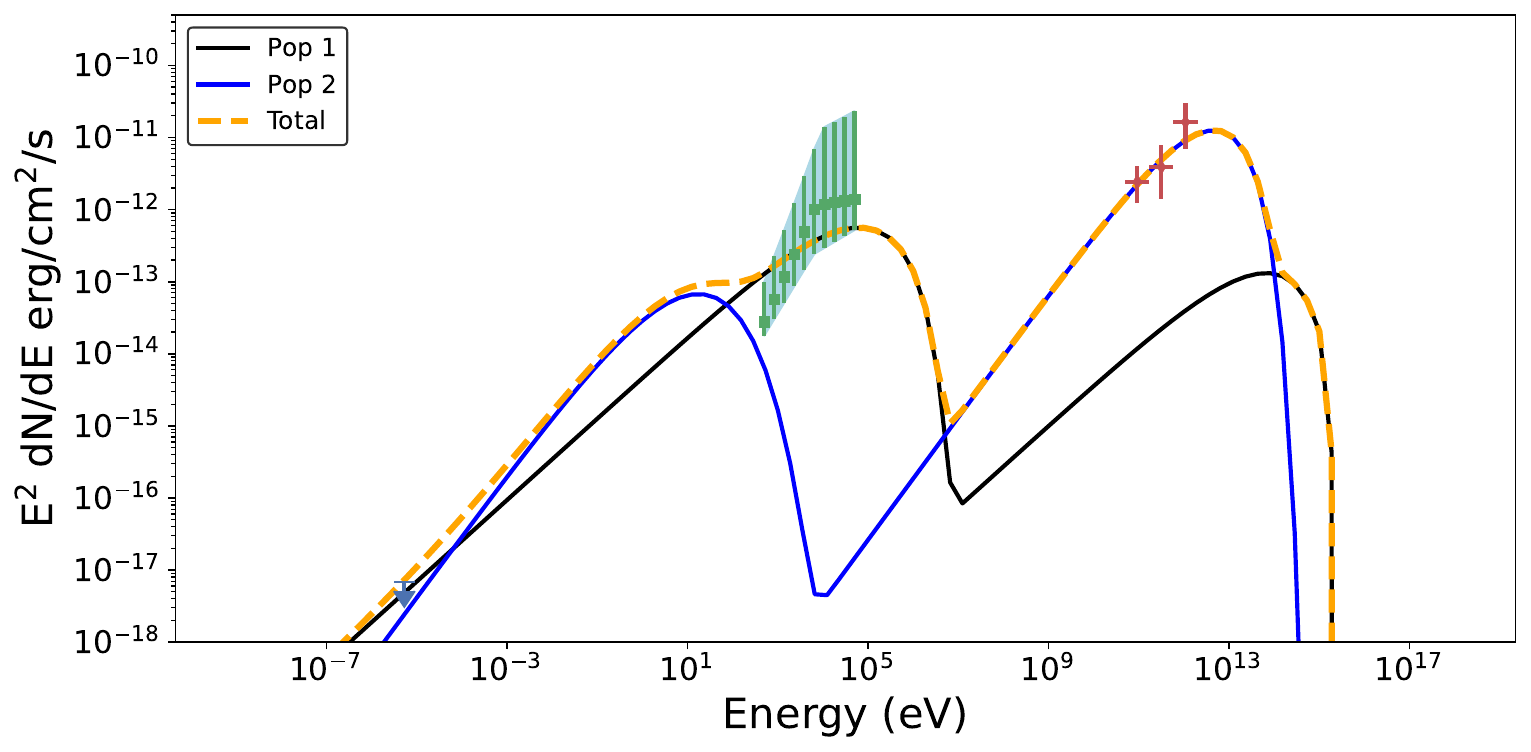}
\end{minipage}
\begin{minipage}{0.49\textwidth} 
\centering
\includegraphics[width=1.0\linewidth]{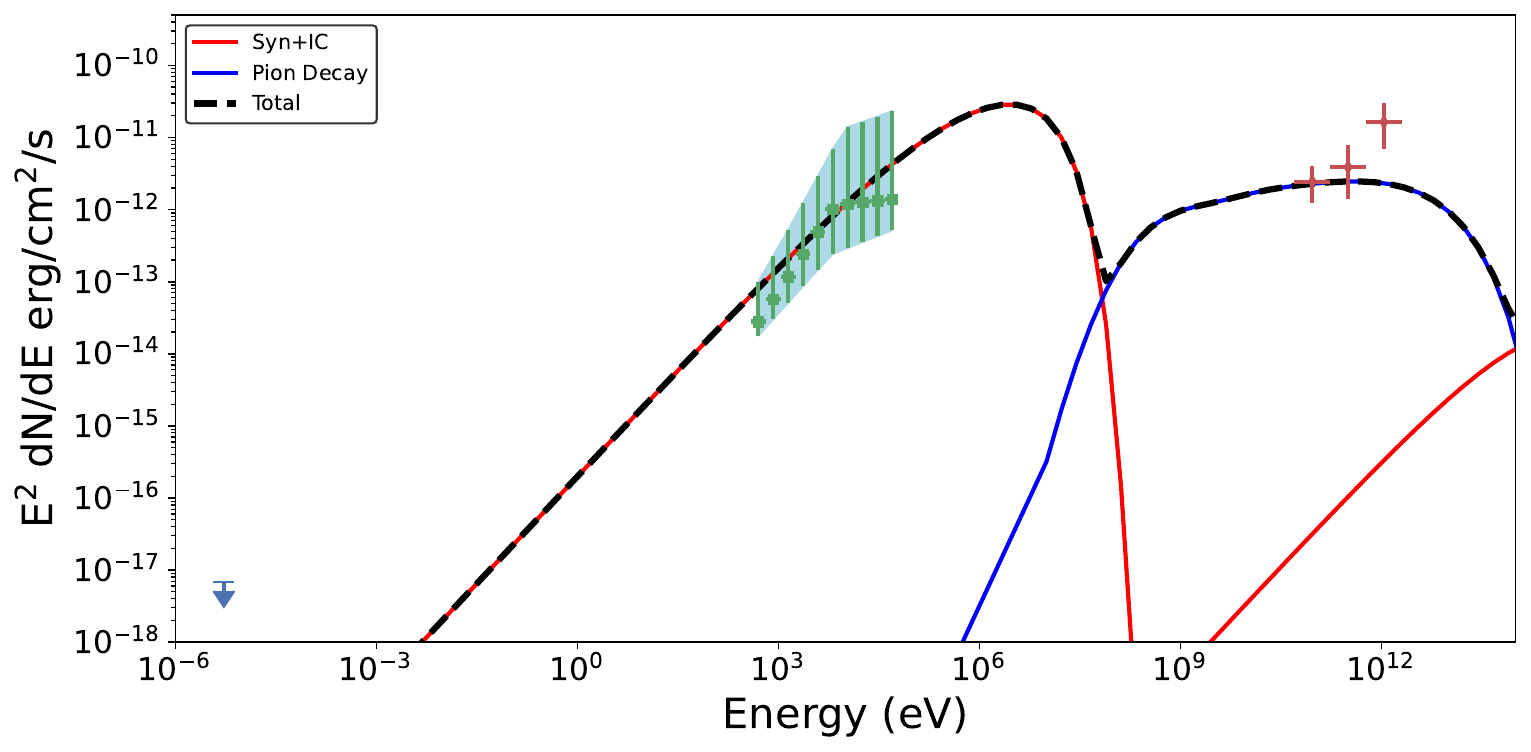}
\end{minipage}
\caption{Broadband representations for the radio, X-ray, and 2FHL $\gamma$-ray data combination. {\it Left:} Best-fit broadband model for the leptonic scenario. Radio data is taken from \citet{heywood2022}, X-ray data is from this work, and Fermi--LAT data from \citet{ackermann16,ajello17}.
{\it Right:} Best-fit broadband model for the lepto-hadronic scenario.}\label{fig:naimamodels_2fhl}
\end{figure*}

\begin{figure*}[t!]
\begin{minipage}{0.49\textwidth} 
\centering
\includegraphics[width=1.0\linewidth]{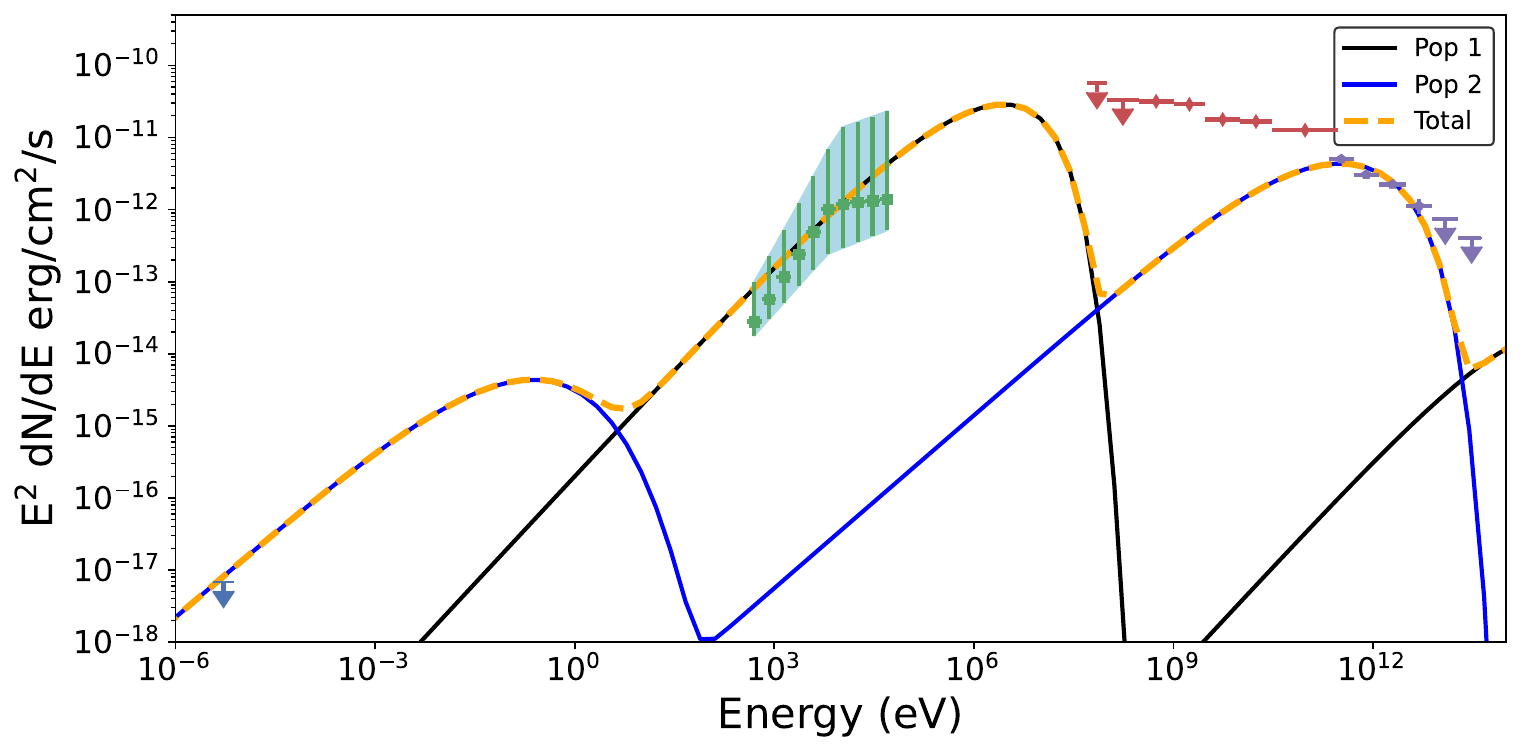}
\end{minipage}
\begin{minipage}{0.49\textwidth} 
\centering
\includegraphics[width=1.0\linewidth]{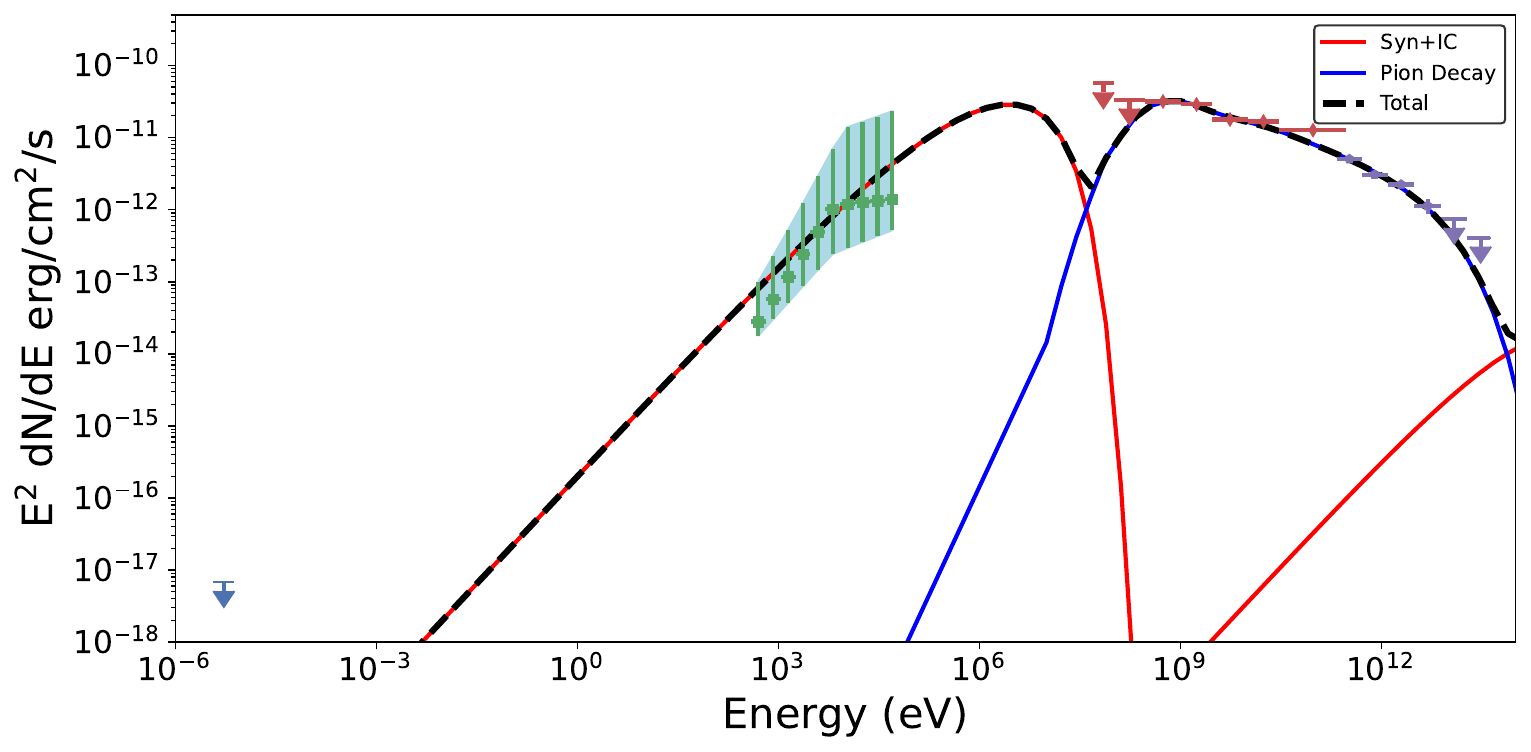}
\end{minipage}
\caption{Broadband representations for the radio, X-ray, 4FGL and \hess\ $\gamma$-ray data combination. {\it Left:} Best-fit broadband model for the leptonic scenario. Radio data is taken from \citet{heywood2022}, X-ray data is from Section~\ref{sec:fit_joint}, Fermi--LAT data from \citet{abdollahi22}, and H.E.S.S. from \citet{hess18}.
{\it Right:} Best-fit broadband model for the lepto-hadronic scenario.}\label{fig:naimamodels}
\end{figure*}

In order to explore a hadronic scenario to the broadband data, we fit a pion decay emission model to characterize the GeV and TeV data together combined with the previously derived leptonic component to characterize the radio and X-ray. As shown in the right panels of Figures~\ref{fig:naimamodels} and \ref{fig:naimamodels_2fhl}, both data combinations are adequately characterized in the lepto-hadronic scenario, and in particular we are now able to efficiently describe the SED including the 4FGL data, something that could not be done with a purely leptonic model.

We note that the fits to both datasets require very large target densities: more specifically, $n_{\rm H} = 5000$\,cm$^{-3}$ in the 4FGL+\hess\ dataset and $n_{\rm H} = 400$\,cm$^{-3}$ in the 2FHL dataset. The high $n_{\rm H}$ values are a requirement if the total proton energy $W_{\rm p}$ is not to exceed the total energy expected to be released from a typical supernova (SN) explosion $E_{\rm SN} = 10^{51}\,$erg. Even for the highest estimate $n_{\rm H} = 5000$\,cm$^{-3}$ which characterizes the 4FGL+\hess\ data, $W_{\rm p} \sim 6 \times 10^{50}$\,erg challenges a hadronic $\gamma$-ray origin. However, both $n_{\rm H}$ and $W_{\rm p}$ may be smaller if there is a significant contribution from non-thermal bremsstrahlung emission (true for both data combinations). A similar representation is presented in \citet{bamba09}, showing that non-thermal bremsstrahlung radiation can dominate the high-energy emission.  

In summary, the SED fitting analysis shows that the two-leptonic model efficiently describes the radio, X-ray, and 2FHL $\gamma$-ray data combination, and in particular provides the most accurate modelization of the 2FHL $\gamma$-ray spectrum. 
On the other hand, the lepto-hadronic representation more efficiently describes the radio, X-ray, and 4FGL+\hess\ extended $\gamma$-ray emission, as shown in the right panel of Figure~\ref{fig:naimamodels}, under the condition that the ambient particle density is high, $n_{\rm H} = 5000$\,cm$^{-3}$. We discuss the implications for the best-fit models in more detail in the following section. 

\begingroup
\begin{table*}[ht!]
\centering
\begin{tabular}{| c | c c |}
\hline
\ 4FGL+\hess\ & Two-Leptonic & Lepto-hadronic \\
\hline
\  $W_{\rm e}$ or $W_{\rm p}^{a}$ (erg) & $4.5 \times 10^{44} \, 1.6 \times 10^{48}$ & $W_{\rm e} = 4.5 \times 10^{44}$\, $W_{\rm p} = 6.2 \times 10^{50}$ \\
\  $n_{\rm H}$ (cm$^{-3}$) & $- \, -$ & $-$\, $5000$ \\
\ $\alpha_{\rm e}$ or $\alpha_{\rm p}^{b}$ & 1.0 \, 1.4 & $\alpha_{\rm e} = 1.0 $\, $\alpha_{\rm p} = 2.4$ \\
\ $E_{e,c}$ or $E_{p,c}^{c}$ (TeV) & 794 \, 3.63 & $E_{e,c} = 794$ \, $E_{p,c} = 71.0$\\
\ $B^{d}$ ($\mu$G) & 37.7 \, 0.145 & 37.7  \\
\hline
\ 2FHL & Two-Leptonic & Lepto-hadronic \\
\hline
\  $W_{\rm e}$ or $W_{\rm p}^{a}$ (erg) & $1.5 \times 10^{46} \, 1.1 \times 10^{48}$ & $W_{\rm e} = 4.5 \times 10^{44}$\, $W_{\rm p} = 2.1 \times 10^{48}$ \\
\  $n_{\rm H}$ (cm$^{-3}$) & $- \, -$ & $-$\, $400$ \\
\ $\alpha_{\rm e}$ or $\alpha_{\rm p}^{b}$ & 1.9 \, 1.3 & $\alpha_{\rm e} = 1.0 $\, $\alpha_{\rm p} = 1.9$ \\
\ $E_{e,c}$ or $E_{p,c}^{c}$ (TeV) & 617 \, 22.4 & $E_{e,c} = 794$ \, $E_{p,c} = 71.0$\\
\ $B^{d}$ ($\mu$G) & 3.0 \, 0.25 & 37.7  \\
\hline
\end{tabular}
\caption{Summary of the best-fit model parameters for the models displayed in Figures~\ref{fig:naimamodels} (4FGL+\hess) and \ref{fig:naimamodels_2fhl} (2FHL). The maximum particle energy is fixed to 2\,PeV in all cases. \footnotesize{$^a$ The total particle energy for electrons $W_{\rm e}$ or protons $W_{\rm p}$, $^b$ the index for electrons $\alpha_{\rm e}$ or protons $\alpha_{\rm p}$, $^c$ the cutoff energy for electrons or protons.}
} 
\label{tab:naima}
\end{table*}
\endgroup

\subsection{Model Interpretation}
In the previous sections, we attempted to determine the most likely origin of the unique and puzzling high-energy source 2FHL~J1745.1--3035 through detailed broadband modeling.
We tested two data combinations assuming two energetic scenarios: 1) the 4FGL+\hess\ extended $\gamma$-ray data is related to the point-like X-ray source or 2) the point-like 2FHL $\gamma$-ray data is related to the point-like X-ray source.
In the previous sections, we determined that the most favorable models correspond to the lepto-hadronic model for the 4FGL+\hess\ dataset, while a two-leptonic model can better characterize the 2FHL dataset. 

The lepto-hadronic model accurately characterizes the source responsible for the 4FGL+\hess\ observations, requiring a large target proton density $n_{\rm H} = 5000$\,cm$^{-3}$. If the distance assumed to the source corresponds to the one reported in  \citet[][]{bamba09}, which is, $d = 8.5\,$kpc, then 2FHL~J1745.1--3035 is likely located within the Galactic Center and thus among dense molecular material. In fact, HESS~J1745--303 is spatially coincident with a molecular cloud \citep{aharonian08} consistent with the estimate $n_{\rm H} = 5000$\,cm$^{-3}$, making this scenario plausible. 

The two-leptonic model best characterizes the emission responsible for 2FHL observations and infers a magnetic field strength that is low compared to the average value for the ISM \citep[$\sim 2\,\mu$G,][]{han1994}. If we assume that all of the considered data are associated in some way, a depicted scenario is one where the extended $\gamma$-ray emission (4FGL+\hess) represents an older, diffuse, and softer spectral component whereas the point-like $\gamma$-ray emission (2FHL) represents a younger, more compact, and harder spectral component. This type of energy-dependent morphology is often observed in evolved PWNe \citep[e.g.,][]{aha2006,hess2012}, but could also be explained by an energetic distant SNR where the highest-energy particles have already escaped and are diffusing into the ambient medium \citep[e.g.,][]{eagle20}. Another possible explanation includes both PWN and SNR contributions, whether they are components of the same system or not. We remark that the models do not consider any electron-to-proton ratio, acknowledging the possibility the emission may be coming from separate, unrelated sources. 

A final scenario that would be compatible with the hard X-ray spectrum we measure is a compact accreting binary system where the compact object, either a neutron star or black hole, is accreting material from a stellar companion. SS~433 represents the most well-known example of such a system and is detected from radio to TeV $\gamma$-rays with extended emission originating from the outflowing jets of the accreting compact object \citep{ss433hawc2018}. Accreting compact binary systems like SS~433 can be extremely luminous in the X-ray band ($\gtrsim 10^{39}$\, erg s$^{-1}$), indicating the the total jet power is much higher \citep{sudoh2020}. The X-ray spectrum is often modeled as synchrotron emission as in the case for SS~433 \citep{safiharb2022}, and can dominate over the Inverse Compton emission if the magnetic field strength is sufficiently high. Lastly, the X-ray emission of these systems can be variable on the timescales of days to hours. The lack of both a clear optical counterpart and of significant X-ray variability overt time-scales of hours (i.e., within each observation) are however two elements that make this scenario less likely with respect to the compact PWN one. 

While the models presented above provide the best representations to the broadband data, there are other limitations to be considered. The most uncertain assumption is the distance used for the modeling, which we assumed to be the same distance reported in \citet{bamba09} for the \hess\ source, which is $d = 8.5\,$kpc. It is possible the source generating the observed emission is closer, which would enable lower target densities and lower total particle energies. Secondly, other particle distributions and hence properties are possible. 
An alternative broadband representation includes a non-thermal bremsstrahlung component which yields reasonable results \citep{bamba09}. Finally, we note that a single population (leptonic or hadronic) cannot explain the broadband data without far exceeding $10^{51}\,$erg in total particle energies at a distance of 8.5\,kpc.

We remark that 2FHL~J1745.1--3035 is unlikely to be of extragalactic origin based on the source selection from the 2FHL catalog, as discussed in Section~\ref{sec:intro}. The spectral index for this source at energies above 50\,GeV is much harder than the average spectral index observed for blazars above 50\,GeV. Furthermore, 2FHL~J1745.1--3035 is located along the Galactic plane and is potentially associated with significant extended GeV and TeV emission (4FGL and \hess\ counterparts). 
Based on this evidence, we conclude that the most plausible scenario for 2FHL~J1745.1--3035 is to be the descendant of a Galactic supernova explosion such as a PWN, SNR, or neutron star or black hole with a stellar companion, but current observations and the presented models prevent us from firmly establishing an origin. 

\section{Summary and conclusions}\label{sec:conclusions}
In this paper, we studied the X- and $\gamma$-ray properties of 2FHL~J1745.1-3035, a very high energy Galactic emitter originally detected by the \lat\ at energies $>$50\,GeV. 2FHL~J1745.1-3035 has a very hard $\gamma$-ray spectrum, with photon index above 50\,GeV $\Gamma_\gamma$=1.2$\pm$0.4, and is found to be a TeV-emitter by the \lat. 
A multi-observatory X-ray campaign with \xmm, \cha\ and \nus\ allowed us to reliably locate the counterpart of 2FHL~J1745.1-3035 and study its properties. The source broadband X-ray spectrum is best-fitted with a broken power law model with break energy E$_{\rm b}$$\sim$7\,keV. The X-ray spectrum is very hard before the break, having photon index $\Gamma_{\rm X,1}$=0.6$\pm$0.4, while it becomes significantly softer at higher energies, with $\Gamma_{\rm X,2}$=1.9$_{-0.2}^{+0.6}$. The X-ray source is compact, with no evidence for extension in the \xmm\ and \nus\ images. However, the subarcsecond \cha\ angular resolution allowed us to detect significantly extended emission up to a scale of $\sim$5$^{\prime\prime}$.

We report below the main results of our analysis. 

\begin{enumerate}
\item The X-ray analysis favors a compact PWN origin for 2FHL~J1745.1-3035: if this is confirmed, it would be one of the hardest PWNe ever detected in the X-rays and the hardest ever detected in the $\gamma$-ray \citep[see Figure~5 in][]{kargaltsev13}. Perhaps the only other Galactic source observed to have such a hard X-ray spectrum is the Crab \citep[e.g.,][]{madsen15,zanin2017}, where the spectrum is associated with synchrotron radiation as in the case developed here for 2FHL~J1745.1--3035.
\item The multi-band observational evidence, and in particular the X-ray one, could be linked to the presence of a young nebula ($t\lesssim$10$^3$ yrs). In fact, both the cut-off and the normalization of the synchrotron spectral curve depend on the magnetic field strength, and the final emission spectrum depends on a combination of the injected particle spectrum, the losses sustained by the population of previously injected particles and the system evolution. Due to these effects, the peak of the synchrotron curve drops in time from roughly 1 MeV at very early times to 1-10 keV when the reverse shock finally crushes the nebula \citep[see, e.g.,][]{gelfand09}. 
At the present day, \lat\ has detected almost only old PWNe \citep[e.g.,][]{acero13}, thus making 2FHL~J1745.1-3035 uniquely interesting. However, we also note that the TeV to X-ray luminosity ratio for this source is $\sim 5$ and is expected to increase with time \citep{mattana09,kargaltsev13}. Some of the youngest PWNe/SNRs have ratios $\lesssim 2$ while older systems have ratios larger than 2 \citep{yamazaki2006}, but the value can vary depending on the system and environment conditions. 
\item  The PWN scenario is also supported by the lack of short-term variability in the single-epoch observations, which is expected if a source is indeed a PWN, while variability might be observed in other classes of compact, X-ray emitting Galactic sources, such as X-ray binaries or cataclysmic variable stars. 
We note that no observational evidence for a pulsar is reported in the literature for the 4FGL source: in particular, a blind pulsation search was performed by \citet{hui11} on the first 29 months of \lat\ observations, but no clear evidence for periodicity was detected.
\item The broadband SED fitting we performed supports a scenario where the compact X-ray source is linked to the 2FHL target and is well fitted by a two-leptonic model, while the 4FGL+\hess\ source requires the presence of an hadronic component to be efficiently modeled. These results are consistent with a scenario where the extended $\gamma$-ray emission (4FGL+\hess) represents an older, diffuse, and softer spectral component whereas the point-like $\gamma$-ray emission (2FHL) represents a younger, more compact, and harder spectral component.
\item Given its TeV-emitter nature, 2FHL~J1745.1-3035 represents an ideal follow-up target for the upcoming Cherenkov Telescope Array \citep[CTA][]{acharya13,cta19}. CTA will have a $\sim$2 times better angular resolution and a $\sim$5 times flux sensitivity  than \hess\ at 1\,TeV \citep{cta19}, and is therefore expected to resolve into individual point sources at least a fraction of the diffuse emission detected by \hess. The study of the Galactic Center region is one of the Key Science Programs of the CTA Observatory, and in particular 2FHL~J1745.1-3035 will be covered by the planned 300-hours Extended Survey. Furthermore, recent preparatory studies for the ASTRI (Astrofisica con Specchi a Tecnologia Replicante Italiana) Mini-Array Observatory \citep{vercellone22} suggest that observations at $\sim$100--300\,TeV can provide important information on the emission processes in young, powerful PWNe, and in particular can put strong constraints on the presence (or lack of) a hadronic component.
\end{enumerate}

In summary, all the observational evidence shows 2FHL~J1745.1-3035 is a rare Galactic \lat\ source, having the hardest $\gamma$--ray photon index in the 2FHL sample and an impressively hard X-ray photon index. Combining all of this evidence hints at a scenario where 2FHL~J1745.1-3035 could be a newly detected, young PWN system. 

The unique properties of this target make it an ideal source for follow-up campaigns with facilities looking at the TeV sky, first and foremost CTA. Furthermore, deeper \cha\ observations would allow one to perform a multi-region X-ray spectral fit to obtain a radial profile of the X-ray photon index and flux, which, if confirmed to be a PWN, could be used to test different PWN emission models to estimate the properties such as the age \citep[see, e.g.,][]{tibaldo17}.

\section*{Acknowledgments}
This research has made use of the NuSTAR Data Analysis Software (NuSTARDAS) jointly developed by the ASI Space Science Data Center (SSDC, Italy) and the California Institute of Technology (Caltech, USA). 
This work makes use of Matplotlib \citep{hunter07} and NumPy \citep{harris20}. 
This work has made use of data from the European Space Agency (ESA) mission {\it Gaia} (\url{https://www.cosmos.esa.int/gaia}), processed by the {\it Gaia} Data Processing and Analysis Consortium (DPAC, \url{https://www.cosmos.esa.int/web/gaia/dpac/consortium}). Funding for the DPAC has been provided by national institutions, in particular the institutions participating in the {\it Gaia} Multilateral Agreement.
This publication makes use of data from the UKIRT Infrared Deep Sky Survey (UKIDSS). The UKIDSS project is defined in \citet{lawrence07}. UKIDSS uses the UKIRT Wide Field Camera \citet[WFCAM;][]{casali07}. The photometric system is described in \citet{hewett06}, and the calibration is described in \citet{hodgkin09}. The pipeline processing and science archive are described in \citet{hambly08}. 
We thank Patrick Slane for the useful conversation and suggestions. S.M. acknowledges financial support from the European Union -- Next Generation EU under the project IR0000012 - CTA+ (CUP C53C22000430006), announcement N.3264 on 28/12/2021: ``Rafforzamento e creazione di IR nell’ambito del Piano Nazionale di Ripresa e Resilienza (PNRR)''. 
A.D. is thankful for the support of the Ram{\'o}n y Cajal program from the Spanish MINECO, Proyecto PID2021-126536OA-I00 funded by MCIN / AEI / 10.13039/501100011033, and Proyecto PR44/21‐29915 funded by the Santander Bank and Universidad Complutense de Madrid.

\bibliographystyle{apj}
\bibliography{2FHLJ1745_X-ray_campaign}

\appendix 
\section{Single epoch X-ray spectra}\label{app:single_epoch}
\begin{figure*}
\begin{minipage}{0.45\textwidth} 
 \centering 
 \includegraphics[width=0.75\textwidth,angle=-90]{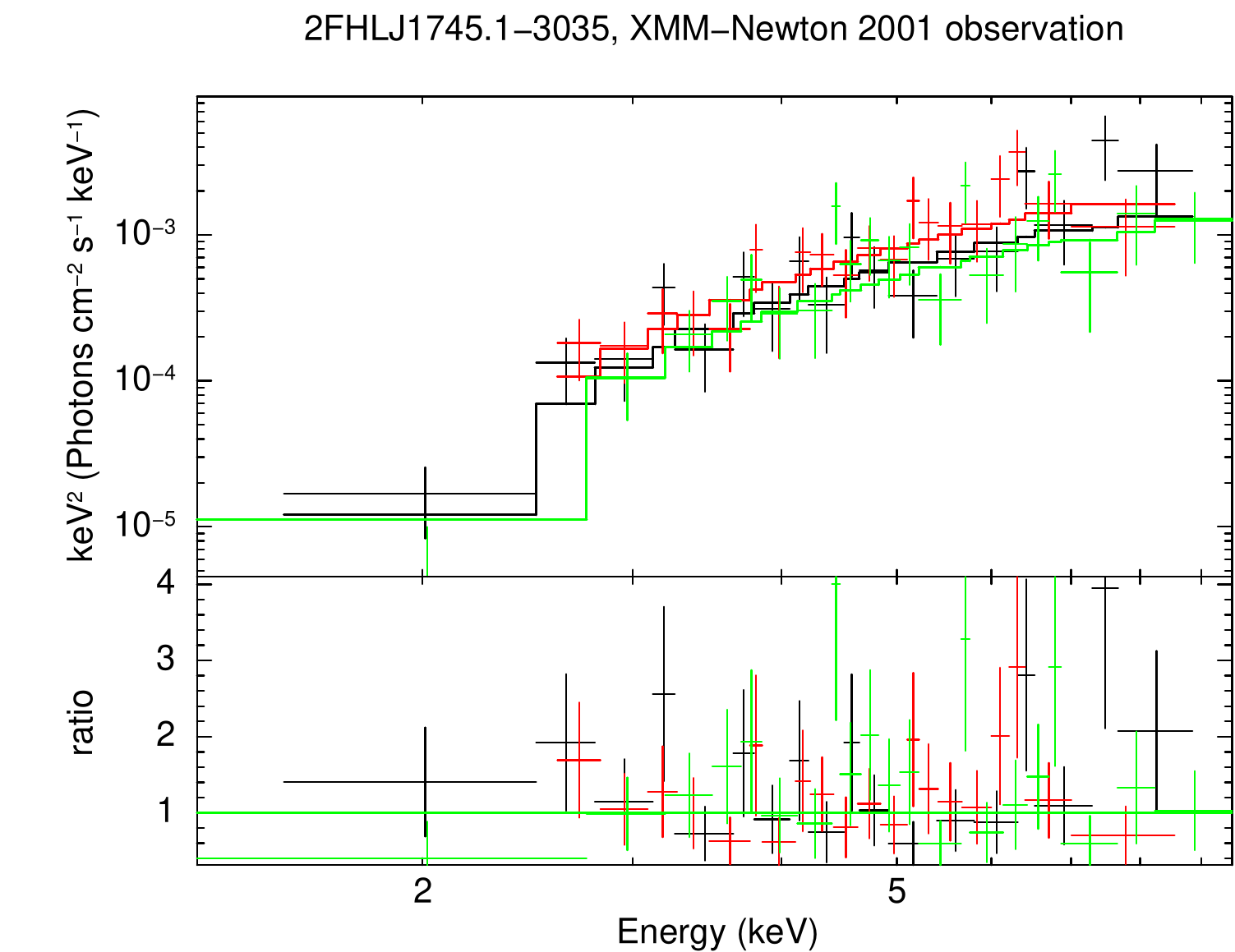} 
 \end{minipage} 
\begin{minipage}{0.45\textwidth} 
 \centering 
 \includegraphics[width=0.75\textwidth,angle=-90]{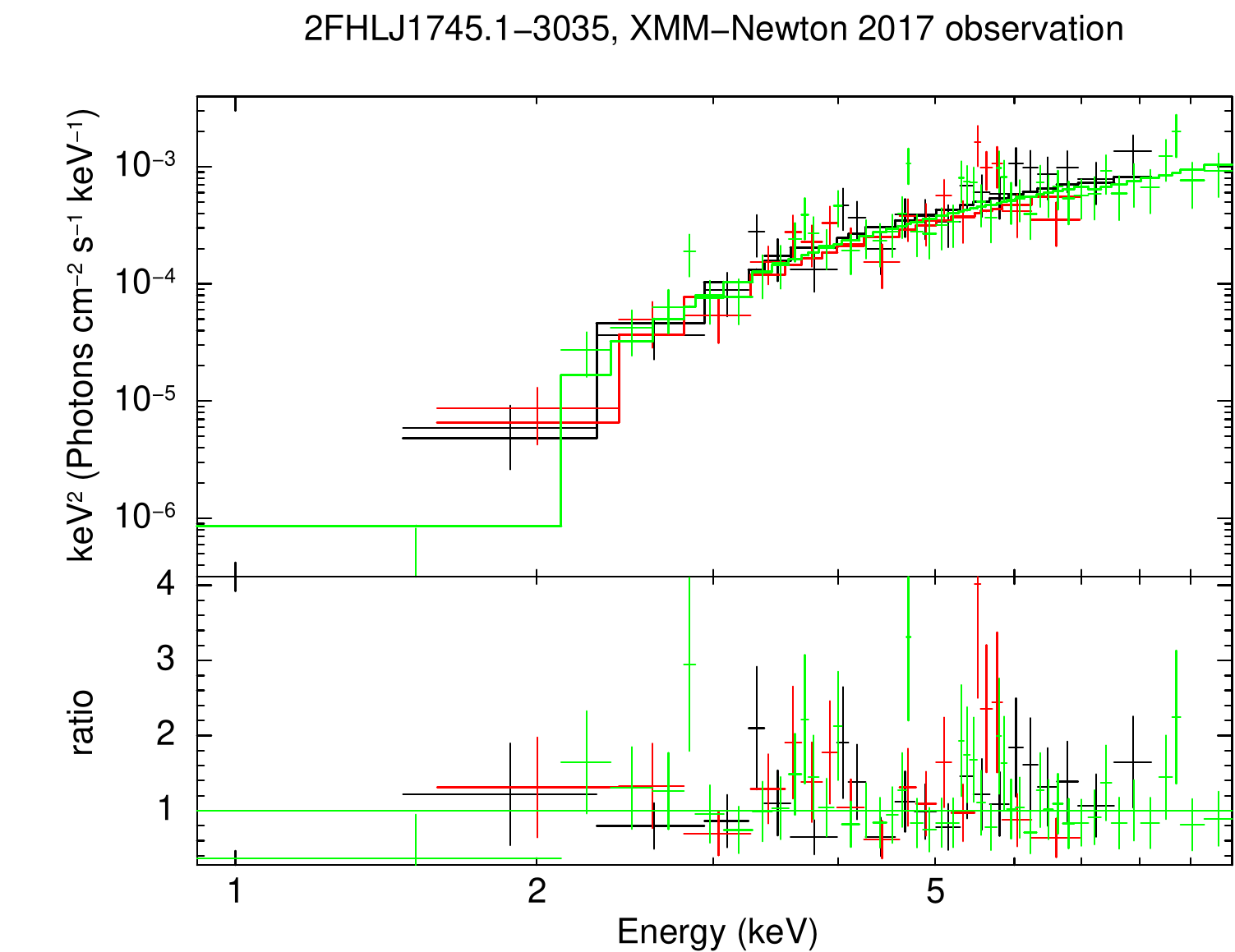}
 \end{minipage}
 \begin{minipage}{0.45\textwidth} 
 \centering 
 \includegraphics[width=0.75\textwidth,angle=-90]{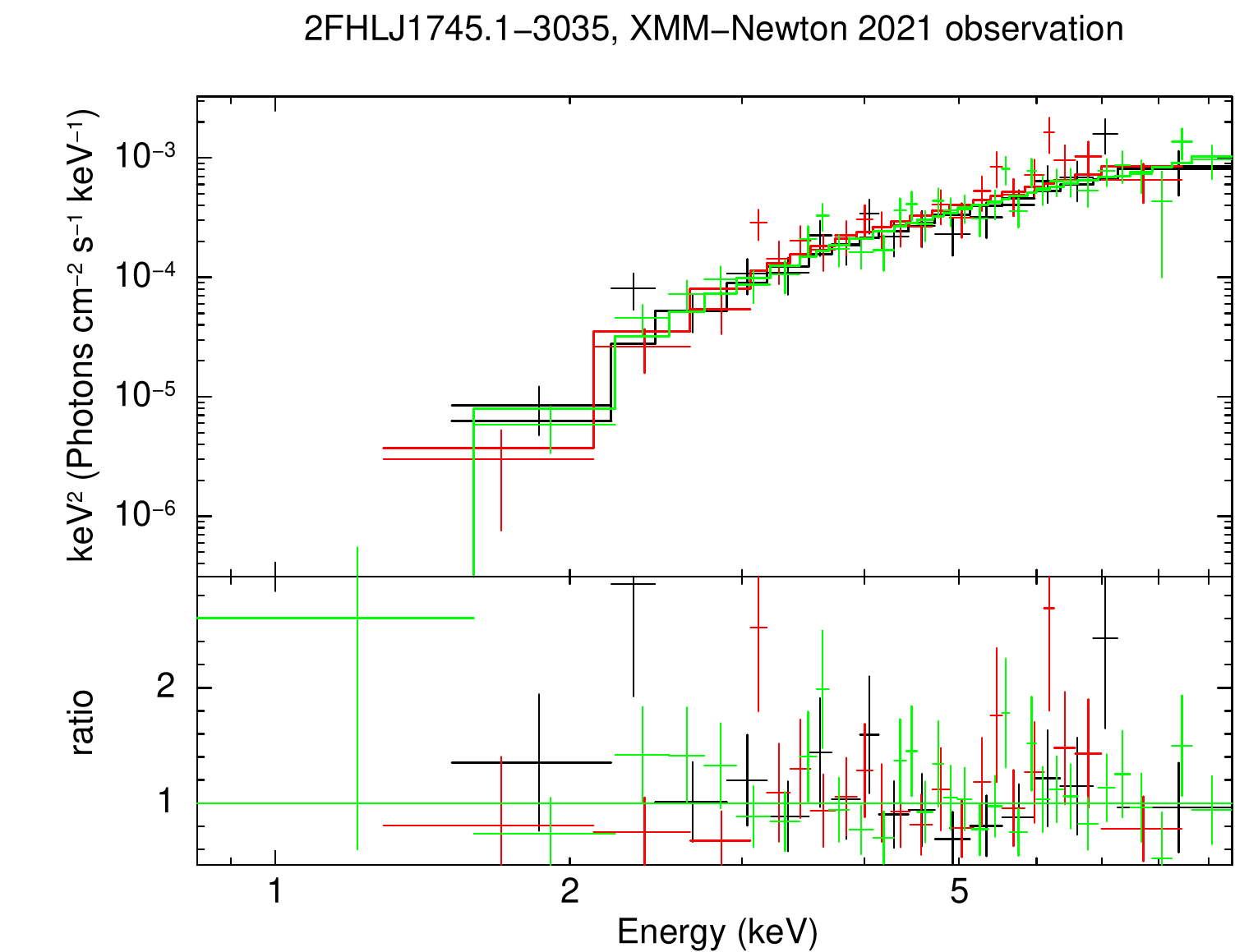}
 \end{minipage}
 \begin{minipage}{0.45\textwidth} 
 \centering 
 \includegraphics[width=0.75\textwidth,angle=-90]{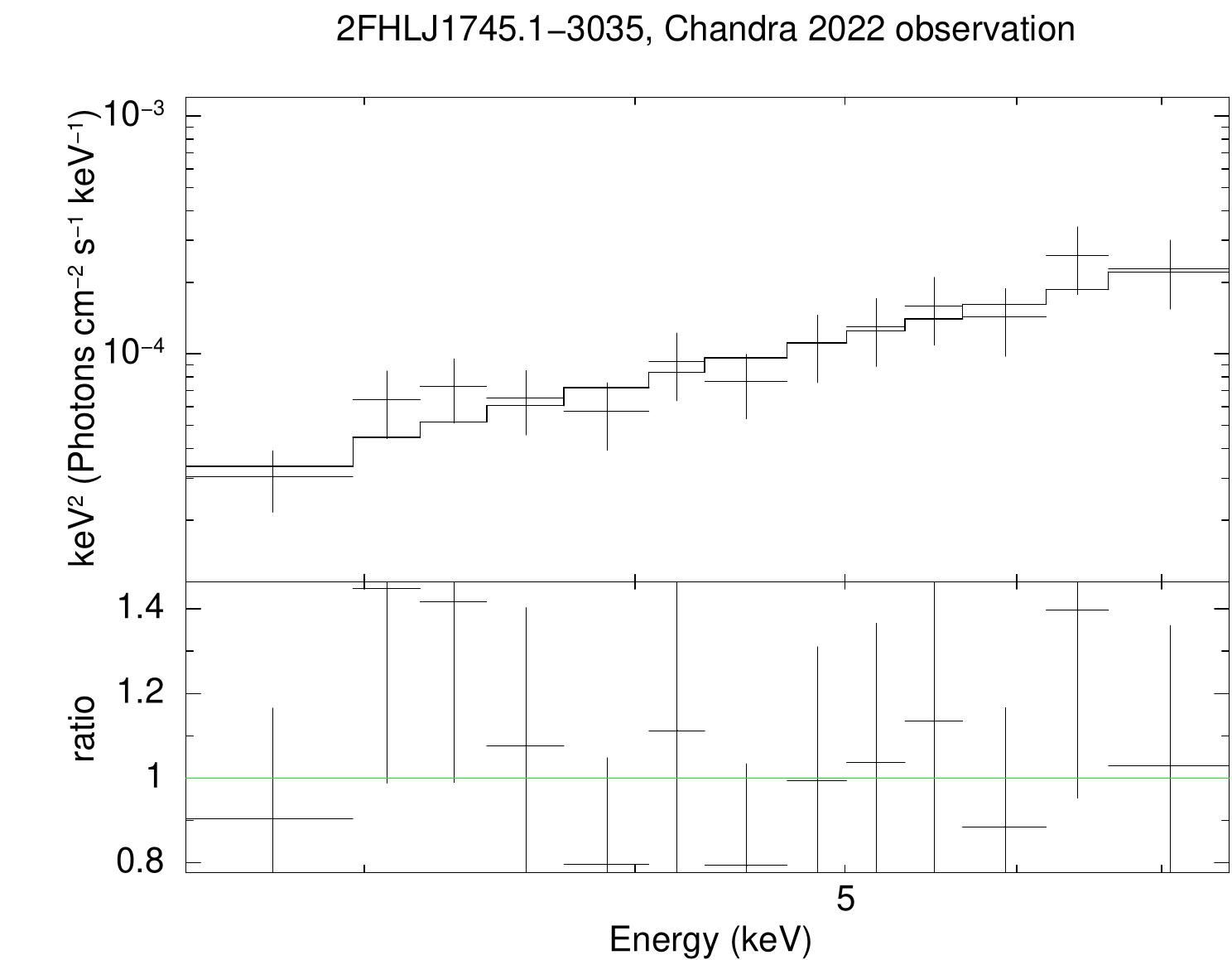}
 \end{minipage}
 \begin{minipage}{0.45\textwidth} 
 \centering 
 \includegraphics[width=0.75\textwidth,angle=-90]{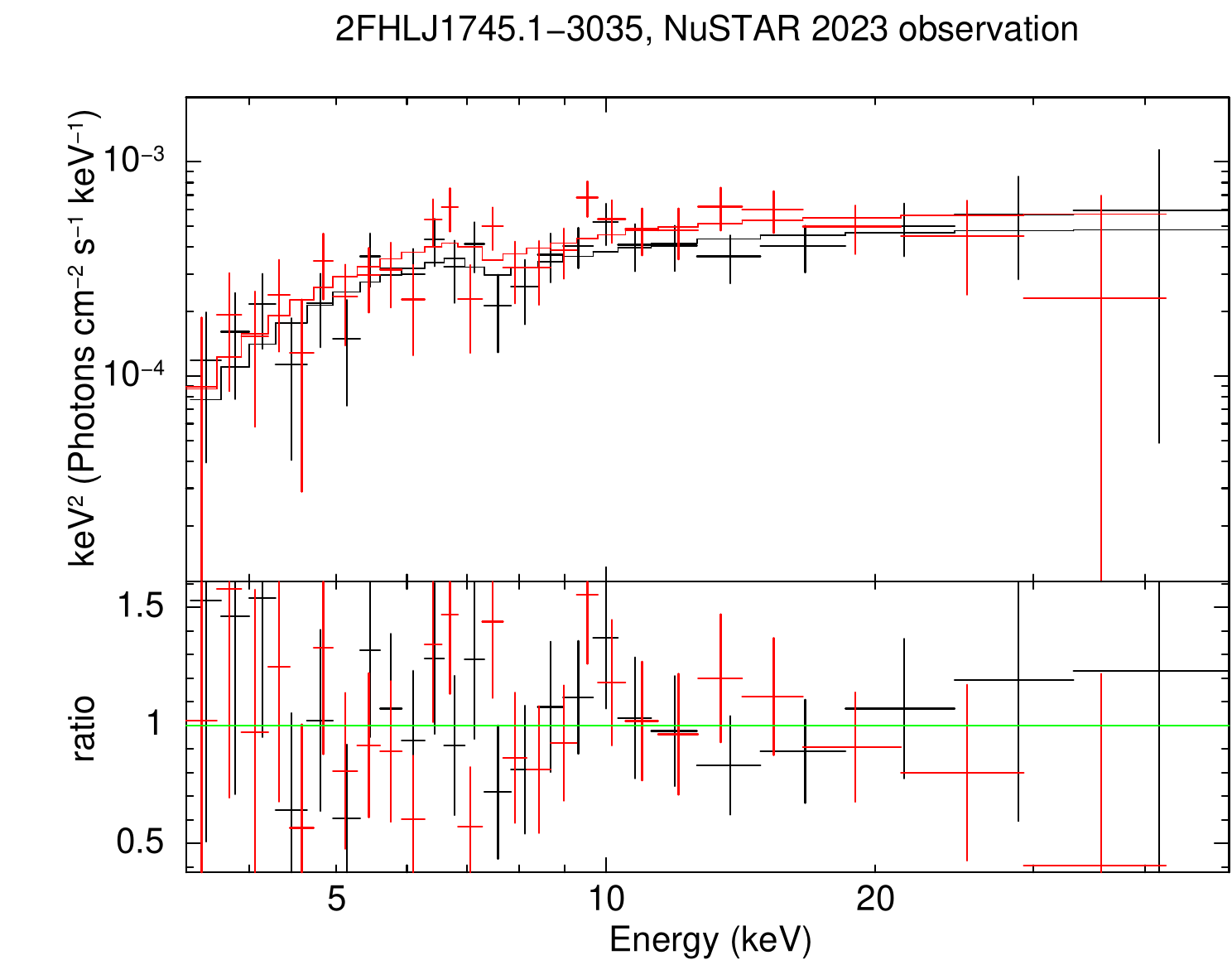}
 \end{minipage}
\caption{\normalsize 
Single epoch X-ray spectra of 2FHL~J1745.1-3035. In the \xmm\ observations, MOS1, MOS2, and pn spectra are plotted in black, red, and green, respectively; in the \nus\ observations, FPMA and FPMB spectra are plotted in black and in red, respectively.  
The best-fit model from an absorbed power law is plotted as a solid line. 
}\label{fig:spectra_single_epoch}
\end{figure*}

We report in this Appendix, in Figure~\ref{fig:spectra_single_epoch}, the single-epoch X-ray spectra of 2FHL~J1745.1-3035. The best-fit models are obtained by fitting the data with an absorbed power law, as discussed in Section~\ref{sec:single_epoch}.

\section{Light curves of the 2017 and 2021 \xmm\ observations}\label{app:light_curves}
\begin{figure*}
\begin{minipage}{0.37\textwidth} 
 \centering 
 \includegraphics[width=1\textwidth]{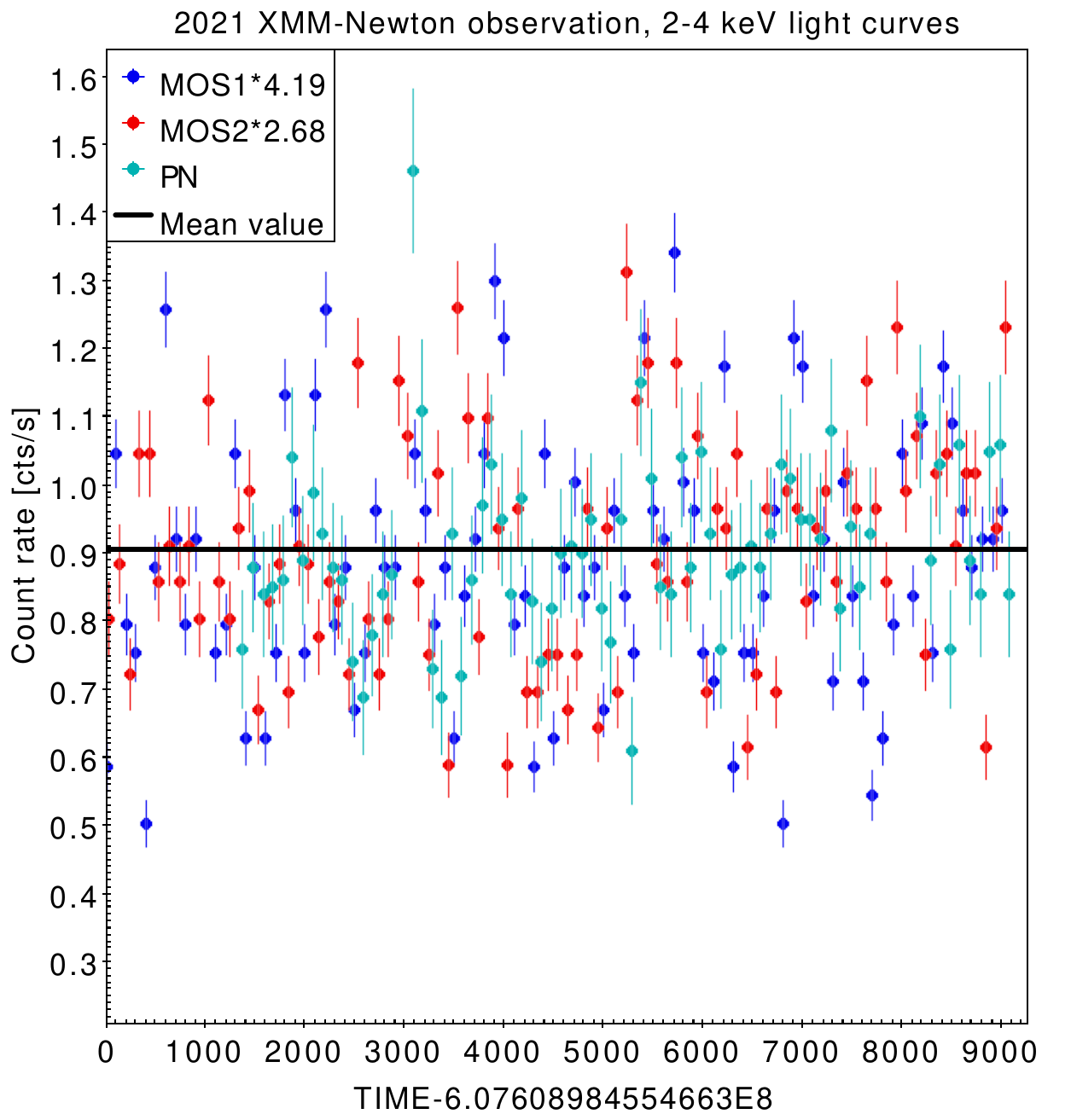} 
 \end{minipage} 
\begin{minipage}{0.37\textwidth} 
 \centering 
 \includegraphics[width=1\textwidth]{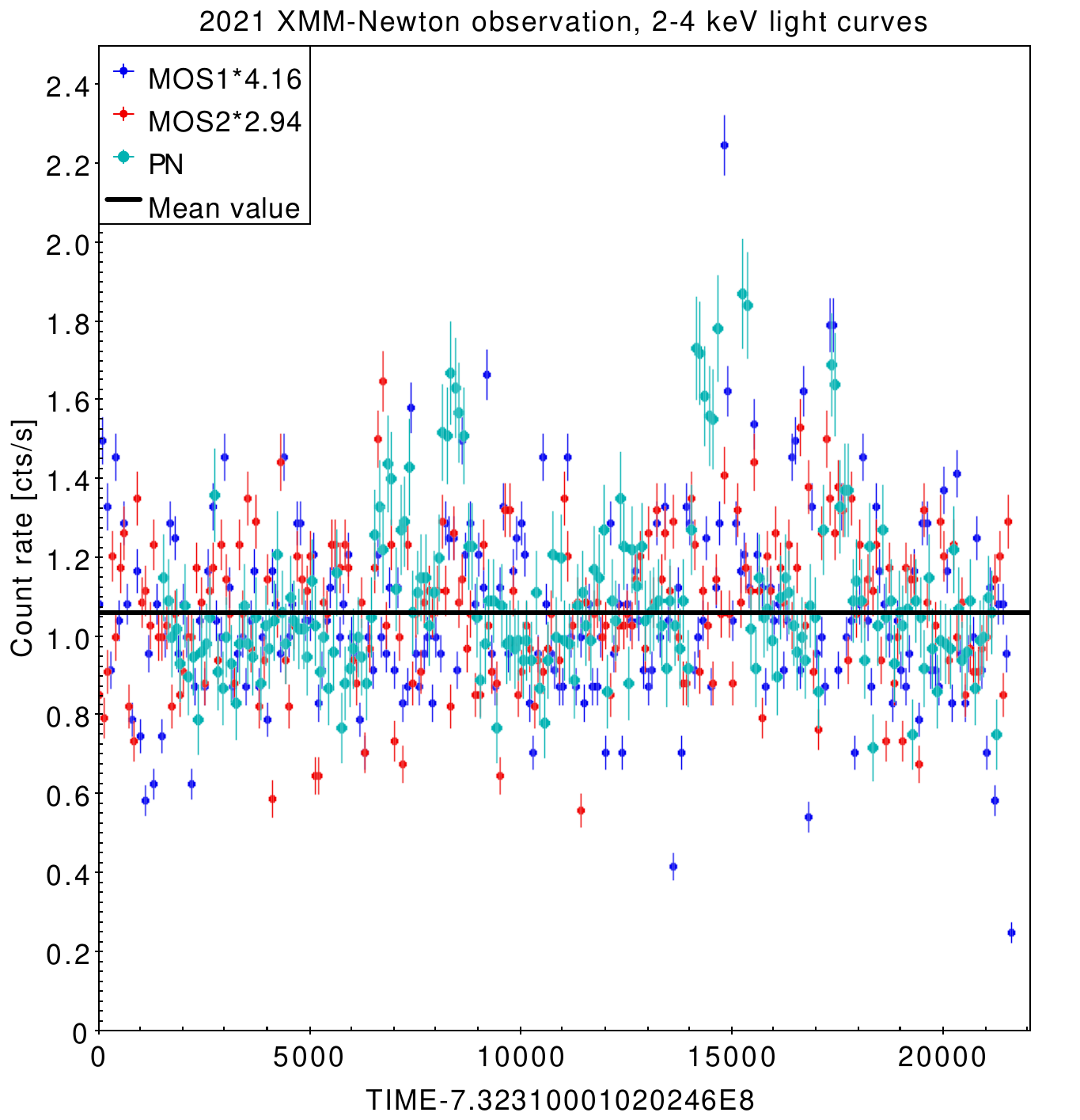} 
 \end{minipage} 
 \begin{minipage}{0.37\textwidth} 
 \centering 
 \includegraphics[width=1\textwidth]{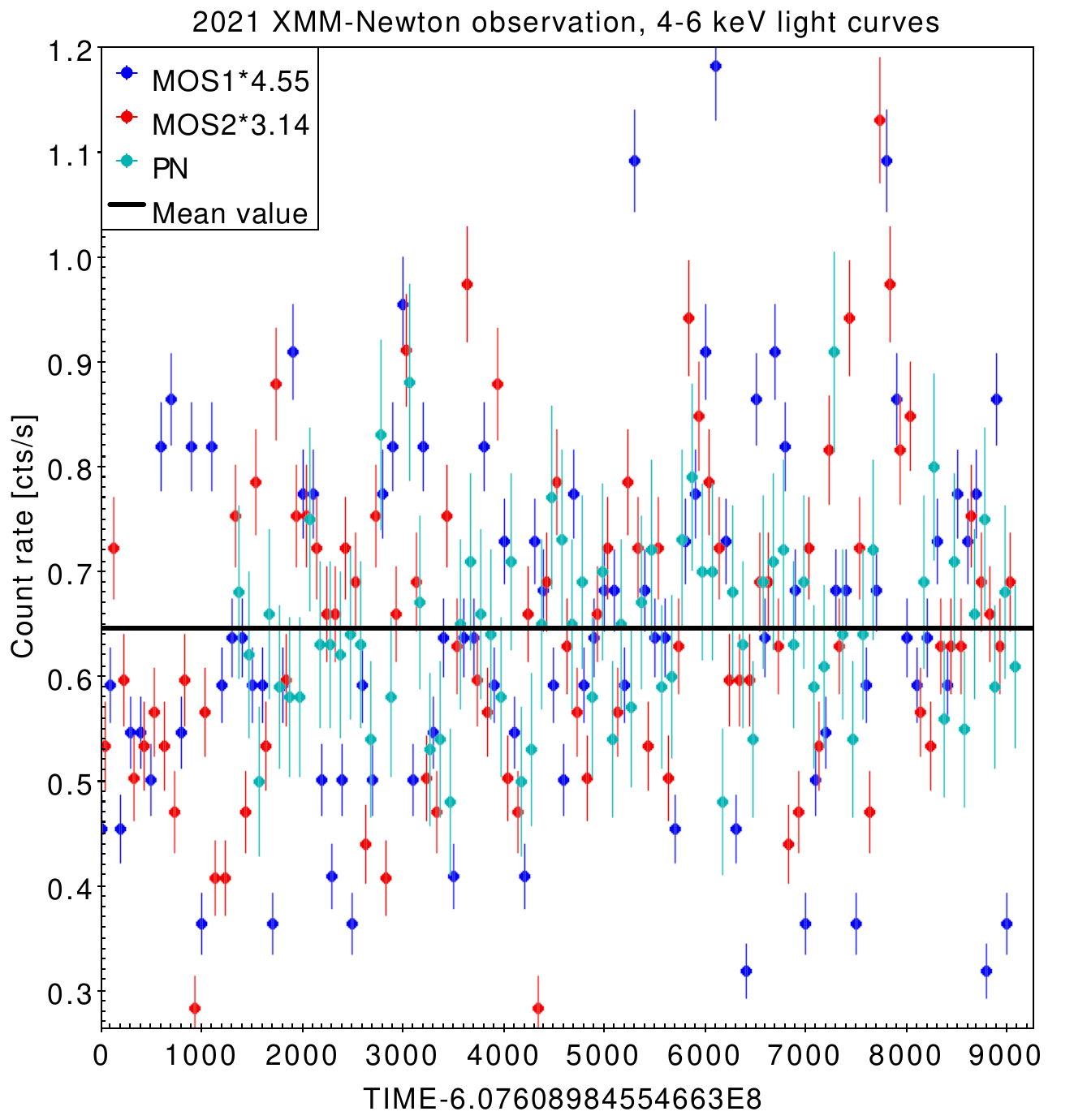} 
 \end{minipage} 
\begin{minipage}{0.37\textwidth} 
 \centering 
 \includegraphics[width=1\textwidth]{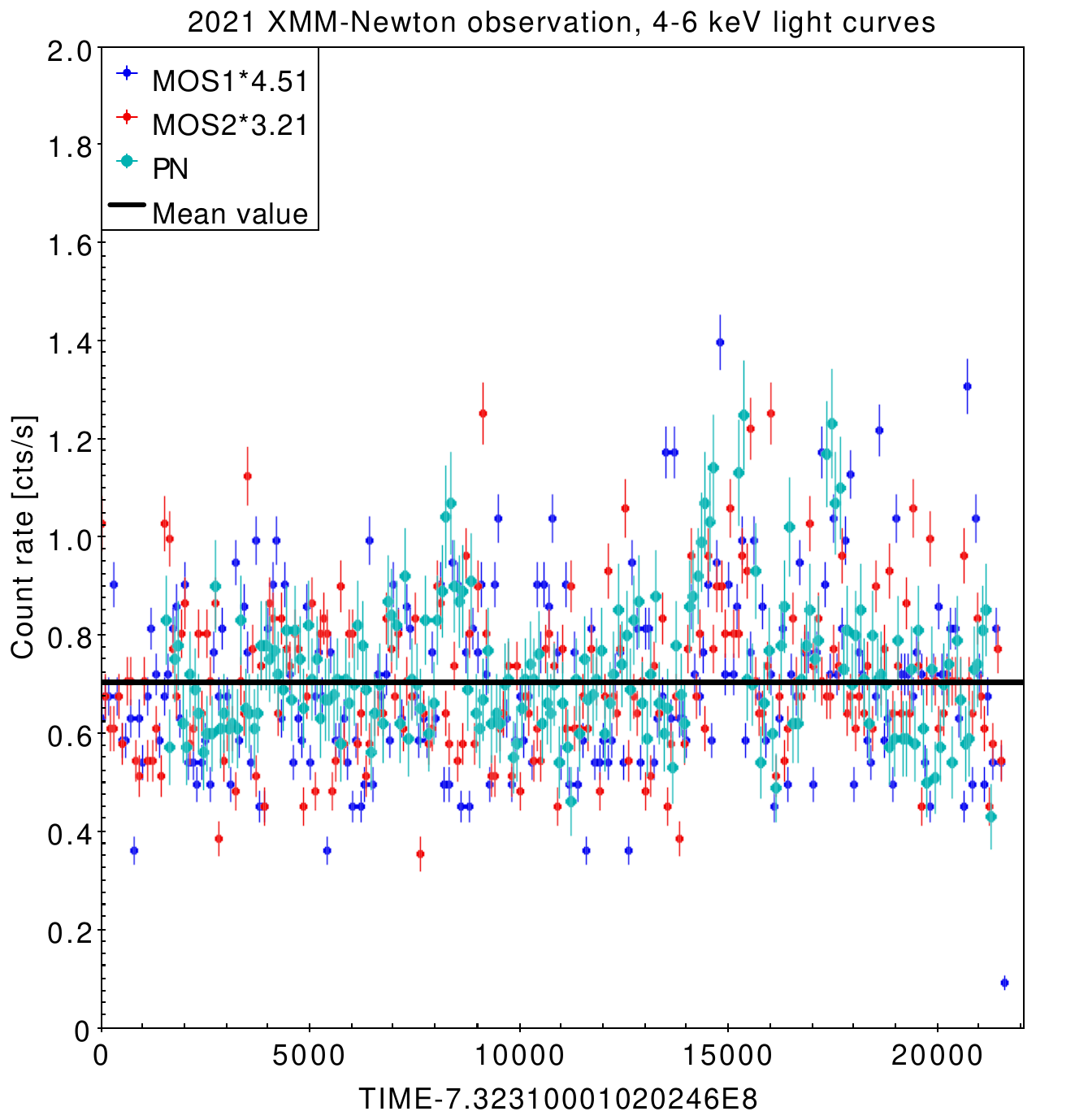} 
 \end{minipage} 
  \begin{minipage}{0.37\textwidth} 
 \centering 
 \includegraphics[width=1\textwidth]{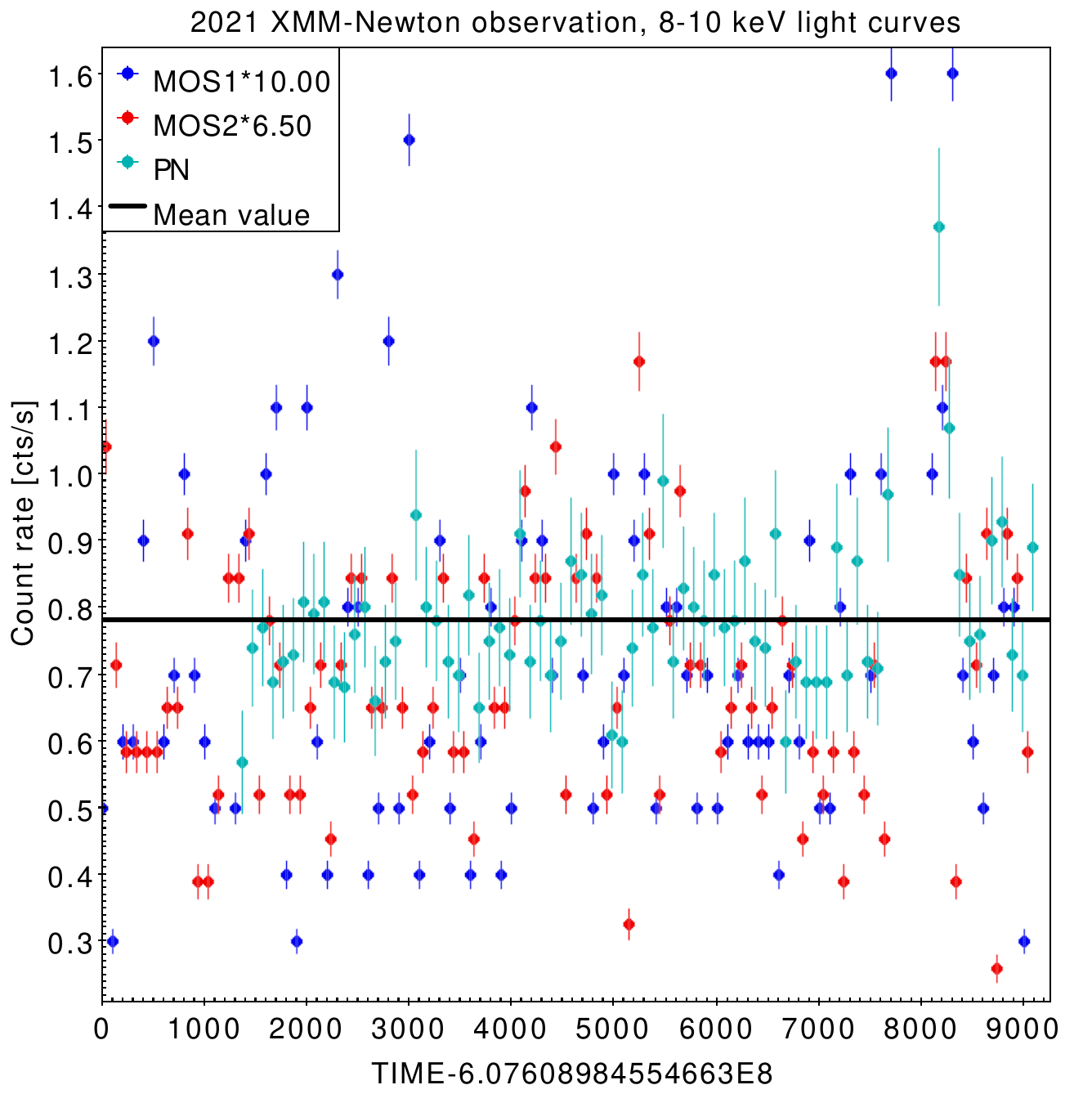} 
 \end{minipage} 
 \hspace{4.2cm}
 \begin{minipage}{0.37\textwidth} 
 \centering 
 \includegraphics[width=1\textwidth]{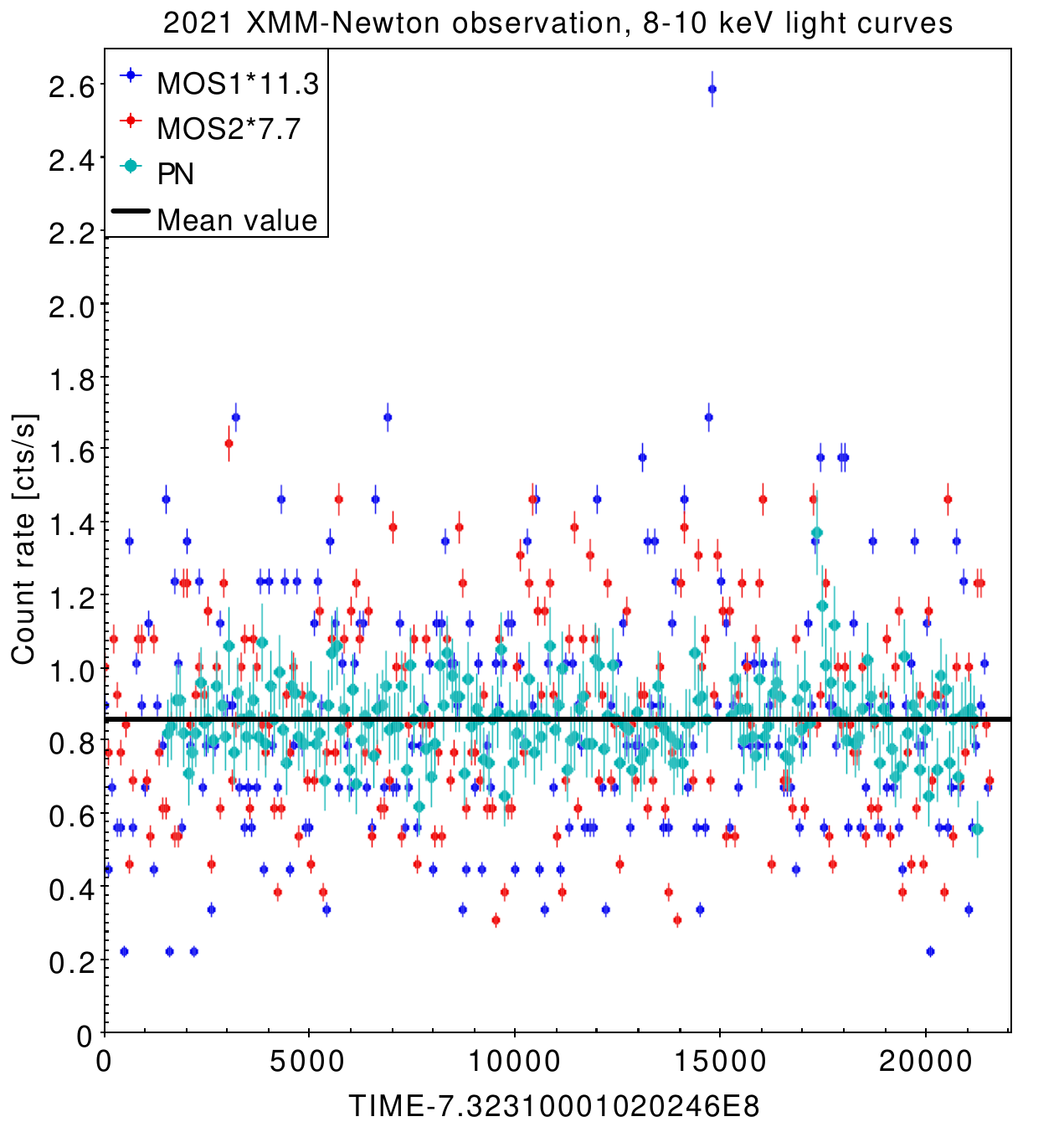} 
 \end{minipage} 
\caption{\normalsize \xmm\ light curves in the 2--4\,keV (top panels), 4--6\,keV (center), and 8--10\,keV band (bottom) for the 2017 (left) and 2021 (right) observations of 4XMM J174507.9-303906. MOS1 data-points are plotted in blue, MOS2 in red, and PN in cyan. The MOS1 and MOS2 light curves have been rescaled so that their average value matches the pn one.
}\label{fig:light_curves}
\end{figure*}

We report in this Appendix, in Figure~\ref{fig:light_curves}, the \xmm\ light curves in the 2--4\,keV, 4--6\,keV, and 6--10\,keV, for the 2017 and 2021 observations of 4XMM J174507.9-303906. We derived these light curves directly from the data: as it can be seen, differently to what is reported in the 4XMM catalog, no clear evidence of cross-instrument variability is visible in the 2021 observations; as for the 2017 observation, significant variability is detected only at energies $>$6\,keV and can be attributed to a strong background flare.

\end{document}